\documentclass[journal]{IEEEtran}

\usepackage{cite}
\usepackage{tabularx,booktabs}
\usepackage{amsmath,amssymb,amsfonts}
\usepackage{graphicx}
\usepackage{algorithm}
\usepackage{algpseudocode}
\usepackage{hyperref}
\usepackage{xcolor}
\usepackage{textcomp}
\usepackage{comment}

\usepackage{multirow}
\usepackage{array}

\usepackage{color, colortbl}

\newcolumntype{L}[1]{>{\raggedright\let\newline\\\arraybackslash\hspace{0pt}}m{#1}}
\newcolumntype{C}[1]{>{\centering\let\newline\\\arraybackslash\hspace{0pt}}m{#1}}
\newcolumntype{R}[1]{>{\raggedleft\let\newline\\\arraybackslash\hspace{0pt}}m{#1}}

\newcolumntype{Y}{>
{\centering\arraybackslash}X}

\def\vsp{\vspace*}
\def\no{\noindent}
\newcommand{\eqq}{$\,=\,$}

\newcommand\sbullet[1][.5]{\mathbin{\vcenter{\hbox{\scalebox{#1}{$\bullet$}}}}}

\DeclareMathAlphabet\mathbfcal{OMS}{cmsy}{b}{n}

\newcommand{\ben}{\begin{eqnarray*}}
	\newcommand{\een}{\end{eqnarray*}}
\newcommand{\be}{\begin{eqnarray}}
\newcommand{\ee}{\end{eqnarray}}

\newcommand{\bv}{\mathbf{v}}
\newcommand{\bx}{\mathbf{x}}
\newcommand{\bhx}{\hat{\mathbf{x}}}

\newcommand{\bz}{\mathbf{z}}
\newcommand{\bhz}{\hat{\mathbf{z}}}

\newcommand{\bw}{\mathbf{w}}

\newcommand{\bB}{\mathbf{B}}
\newcommand{\bQ}{\mathbf{Q}}
\newcommand{\bP}{\mathbf{P}}
\newcommand{\bK}{\mathbf{K}}

\newcommand{\bbN}{\mathbb{N}}
\newcommand{\bbR}{\mathbb{R}}

\newcommand{\ccX}{\mathcal{X}}

\newcommand{\ccN}{\mathcal{N}}

\newcommand{\bcX}{\mathbfcal{X}}

\newcommand{\bet}{\boldsymbol{\eta}}
\newcommand{\bhet}{\hat{\boldsymbol{\eta}}}

\newcommand{\la}{\lambda}
\newcommand{\alp}{\alpha}
\newcommand{\ka}{\kappa}

\newcommand{\LR}{\star}
\newcommand{\TR}{}

\usepackage{acro}
	\DeclareAcronym{wf}{
	short = WF ,
	long  = Wiener Filter ,
}
\DeclareAcronym{fir}{
	short = FIR ,
	long  = Finite Impulse Response ,
}
\DeclareAcronym{wss}{
	short = WSS ,
	long  = Wide Sense Stationary ,
}
\DeclareAcronym{wna}{
	short = WNA ,
	long  = White Noise Acceleration ,
}
\DeclareAcronym{mmse}{
	short = MMSE ,
	long  = Minimum Mean Square Error ,
}
\DeclareAcronym{rmse}{
	short = RMSE ,
	long  = root-mean square error ,
}
\DeclareAcronym{mse}{
	short = MSE ,
	long  = mean square error ,
}
\DeclareAcronym{ssr}{
	short = SSR ,
	long  = state space representation ,
}
\DeclareAcronym{iid}{
	short = i.i.d. ,
	long  = independent and identically distributed ,
}
\DeclareAcronym{pdf}{
	short = PDF ,
	long  = probability density function ,
}
\DeclareAcronym{pdfs}{
	short = PDFs ,
	long  = probability density functions ,
}
\DeclareAcronym{pmf}{
	short = PMF ,
	long  = probability mass function ,
}
\DeclareAcronym{kf}{
	short = KF ,
	long  = Kalman filter ,
}
\DeclareAcronym{ekf}{
	short = EKF ,
	long  = extended Kalman filter ,
}
\DeclareAcronym{ukf}{
	short = UKF ,
	long  = unscented Kalman filter ,
}
\DeclareAcronym{ut}{
	short = UT ,
	long  = unscented transformation ,
}
\DeclareAcronym{pf}{
	short = PF ,
	long  = particle filter ,
}
\DeclareAcronym{sis}{
	short = SIS ,
	long  = Sequential Importance Sampling ,
}
\DeclareAcronym{sir}{
	short = SIR ,
	long  = Sequential Importance Resampling ,
}
\DeclareAcronym{sis-pf}{
	short = SIS-PF ,
	long  = sequential importance sampling - particle filter ,
}
\DeclareAcronym{sir-pf}{
	short = SIR-PF ,
	long  = sequential importance resampling - particle filter ,
}
\DeclareAcronym{mc}{
	short = MC ,
	long  = Monte Carlo ,
}
\DeclareAcronym{mcmc}{
	short = MCMC ,
	long  = Markov chain Monte Carlo ,
}
\DeclareAcronym{smc}{
	short = SMC ,
	long  = Sequential Monte Carlo ,
}
\DeclareAcronym{2d}{
	short = 2D ,
	long  = two dimensional ,
}
\DeclareAcronym{3d}{
	short = 3D ,
	long  = three dimensional ,
}
\DeclareAcronym{snr}{
	short = SNR ,
	long  = signal-to-noise ratio ,
}
\DeclareAcronym{apf}{
	short = APF ,
	long  = auxiliary particle filter ,
}
\DeclareAcronym{awgn}{
	short = AWGN ,
	long  = additive white Gaussian noise ,
}
\DeclareAcronym{tbd}{
	short = TBD ,
	long  = track-before-detect ,
}
\DeclareAcronym{pf-tbd}{
	short = PF-TBD ,
	long  = particle filter based track-before-detect ,
}
\DeclareAcronym{af}{
	short = AF ,
	long  = ambiguity function ,
}
\DeclareAcronym{fov}{
	short = FOV ,
	long  = field of view ,
}
\DeclareAcronym{rf}{
	short = RF ,
	long  = radio frequency ,
}
\DeclareAcronym{eo}{
	short = EO ,
	long  = electro-optical ,
}
\DeclareAcronym{eef}{
	short = EEF ,
	long  = exponentially embedded family ,
}
\DeclareAcronym{kl}{
	short = KL ,
	long  = Kullback-Leibler ,
}
\DeclareAcronym{mh}{
	short = M-H ,
	long  = Metropolis-Hastings ,
}
\DeclareAcronym{rfs}{
	short = RFS ,
	long  = random finite set ,
}
\DeclareAcronym{phdf}{
	short = PHDF ,
	long  = probability hypothesis density filter ,
}
\DeclareAcronym{phdf-pf}{
	short = PHDF-PF ,
	long  = particle implementation of the probability hypothesis density filter,
}
\DeclareAcronym{phd}{
	short = PHD ,
	long  = probability hypothesis density,
}
\DeclareAcronym{pgfl}{
	short = PGFL ,
	long  = probability generating functional ,
}
\DeclareAcronym{gmm}{
	short = GMM ,
	long  = Gaussian mixture model ,
}
\DeclareAcronym{memberf}{
	short = MeMBerF ,
	long  = multi-object multi-Bernoulli filter ,
}
\DeclareAcronym{cbmemberf}{
	short = CBMeMBerF ,
	long  = cardinality balanced multi-object multi-Bernoulli filter ,
}
\DeclareAcronym{cbmemberf-pf}{
	short = CBMeMBerF-PF ,
	long  = particle implementation of the cardinality balanced multi-object multi-Bernoulli filter ,
}
\DeclareAcronym{omat}{
	short = OMAT ,
	long  = optimal mass transfer ,
}
\DeclareAcronym{ospa}{
	short = OSPA ,
	long  = optimal subpattern assignment ,
}
\DeclareAcronym{em}{
	short = EM ,
	long  = Expectation Maximization ,
}
\DeclareAcronym{dp}{
	short = DP ,
	long  = Dirichlet Process ,
}
\DeclareAcronym{hdp}{
	short = HDP ,
	long  = Hierarchical Dirichlet Process ,
}

\DeclareAcronym{crp}{
	short = CRP ,
	long  = Chinese restaurant process ,
}

\DeclareAcronym{tl}{
	short = TL ,
	long  = transfer learning ,
}

\DeclareAcronym{tl-ukf}{
	short = TL-UKF ,
	long  = transfer learning-unscented Kalman filter ,
}

\DeclareAcronym{ckf}{
	short = CKF ,
	long  = cubature Kalman filter ,
}

\DeclareAcronym{tl-ckf}{
	short = TL-CKF ,
	long  = transfer learning-cubature Kalman filter ,
}

\DeclareAcronym{mvf}{
	short = MVF ,
	long  = measurement vector fusion ,
}

\DeclareAcronym{btl}{
	short = BTL ,
	long  = Bayesian transfer learning ,
}

\DeclareAcronym{btlf}{
	short = BTLF ,
	long  = Bayesian transfer learning filter ,
}

\DeclareAcronym{fpd}{
	short = FPD ,
	long  = fully probabilistic design ,
}

\DeclareAcronym{dkf}{
	short = DKF ,
	long  = distributed Kalman filter ,
}

\DeclareAcronym{dtnet}{
	short = DTNet ,
	long  = Detection Tracking Network ,
}

\DeclareAcronym{v2v}{
	short = V2V ,
	long  = vehicle-to-vehicle ,
}

\begin{document}
\title{Object Tracking Incorporating Transfer Learning into Unscented and Cubature Kalman Filters}
\author{
Omar A. Alotaibi, \IEEEmembership{Senior Member,~IEEE}, 
Brian L.\ Mark, \IEEEmembership{Senior Member,~IEEE}, \\
and Mohammad Reza Fasihi, \IEEEmembership{Student Member,~IEEE}
\vspace*{-25pt}
\thanks{A preliminary version of this work was presented in~\cite{Alotaibi-2024-TL_UKF}. 
Omar A.\ Alotaibi is with the Department of Computer Engineering, College of Computer and Information Sciences, King Saud University,
Riyadh 11543, Saudi Arabia (e-mail: oaalotaibi@ksu.edu.sa). 
Brian L.\ Mark and Mohammad Reza Faishi are with the Department of Electrical and Computer Engineering, George Mason University, Fairfax, VA 22030 USA (e-mail: bmark@gmu.edu; mfasihi4@gmu.edu).}}



\maketitle

\begin{abstract}
We present a novel filtering algorithm that employs Bayesian transfer learning to address the challenges posed by mismatched intensity of the noise in a pair of tracking systems, each of which
tracks an object using a nonlinear dynamic system model.  In this setting, 
the primary filter experiences a higher
noise intensity in tracking the object than the source filter. To improve the estimation accuracy of the primary filter, we propose a framework that integrates Bayesian transfer learning into an \ac{ukf} and a \ac{ckf}. The parameters of the predicted observations in the source filter are transferred to the primary filter and used as an additional prior in the
filtering process.  Our simulation results show that the transfer learning approach 
significantly outperforms the conventional isolated \ac{ukf} and \ac{ckf}.
Comparisons to measurement vector fusion are also presented.
\end{abstract}


\acresetall

\section{Introduction}

\IEEEPARstart{T}{racking} models of dynamic systems have been widely used in applications where achieving high  estimation accuracy is critical. Bayesian filters are often used in such applications, although optimal solutions do not exist
for tracking nonlinear models~\cite{Kushner-1967-approximations_Bayesian}. 
The \ac{ukf}, introduced in~\cite{JulieU-1997-UKF}, is a suboptimal Bayesian filter 
for nonlinear model tracking that relies on a Gaussian assumption based on
a set of sigma points drawn using the \ac{ut}. In~\cite{Arasaratnam-2009-CKF}, another approximate filter for nonlinear models, the \ac{ckf}, was proposed, based on the third-degree cubature rule to achieve better accuracy and numerical stability. Although a single tracking system using the \ac{ukf} and \ac{ckf} has been well studied, less work has been done on the application of two separate tracking system scenarios compared to a single tracking system. Moreover, in scenarios of two separated tracking systems, it is often assumed that all tracking systems operate under the same environmental conditions. In practice, this assumption may not always be valid because each tracking system may encounter varying conditions, resulting in variations in the accuracy of the estimation among tracking systems.

In this paper, we consider two separated tracking systems in which both filters track
an object using a \ac{ukf} or \ac{ckf} while experiencing different noise intensities.
We designate one of the tracking filters as the source filter and the other as the primary
filter.  We shall assume that the primary tracking filter experiences higher noise intensity
than the source filter.  To improve the estimation accuracy of the primary filter,
the source filter shares the outcome of the source tracking task instead of the raw data from the source domain.  This is known in the machine learning literature as transfer learning~\cite{Pan-2009-survey_TL}.

Transfer learning has been used to address filtering sparsity problems~\cite{Li-2009_TL_Co_Filtering,Grolman-2016_TL_Co_Filtering_2}. Bayesian models have been used in transfer learning approaches for probabilistic graphical models, such as sharing a Gaussian prior~\cite{Xuan-2021_bayesian_TL}. The incorporation of transfer learning with
Bayesian approaches to improve the performance of the target domain 
is referred to as \ac{btl}~\cite{Karbalayghareh-2018-optimal_TL,Wu-2023-bayesian_TL}. In \cite{Papevz-2022-BTL,Karbalayghareh-2018-TL_regression}, \ac{btl} was used to model the interacting tasks of Gaussian process regression. Moreover, transfer learning has been paired with a Bayesian inference framework to transfer a visual prior learned for online object tracking in challenging environments~\cite{Wang-2012-TL_tracking_image}. In \cite{Foley-2017_fully,Papevz-2019-robust,Papevz-2021-hierarchical}, \ac{btl} was applied in the context
of two separated tracking systems with mismatched noise intensities
in which each tracking system employs a Kalman filter to track
a linear motion model. The source filter shares
its predicted observations to improve the estimation
performance of the primary filter.

Our work extends the \ac{btl} approach referred to as \ac{fpd} in~\cite{Foley-2017_fully} to the case of tracking a {\em nonlinear} model of a dynamical system with  the \ac{ukf} and \ac{ckf}.  In particular,
we adopt the \ac{fpd} variant approach of~\cite{Foley-2017_fully}, in which
both the mean and covariance of the {\em predicted} 
observations of the source filter are
transferred to the primary filter, rather than
the observed raw measurement data~\cite{Pan-2009-survey_TL}. 
The size of the predicted observation data is significantly smaller than the raw,
cluttered observed data from the source domain~\cite{Papevz-2022-BTL}, which are
used in the centralized \ac{mvf} approach~\cite{Willner-1976-MVF} and also in the \ac{dkf} \cite{Olfati-2007_DKF,Talebi-2018_DKF}.

In scenarios where multiple tracking systems are operated by different operators within the same coverage area, each operator utilizes a distinct configuration.  In these decentralized scenarios, \ac{mvf} can not be applied. 
On the other hand, while \ac{dkf} allows for the sharing of estimated states or measurements, it faces several drawbacks compared to the \ac{btlf} approach. First, sharing measurements in a cluttered environment provides significant communication overhead due to the transfer of unwanted data. Second, the tracking process is time-sensitive, as it require sequential updates on per second basis.
This requires transferring shared information between operators within less than a second to be incorporated into the filtering process, which can be impractical. Last but not least, privacy risk challenges of reverse engineering, jamming, or blind spots identification are posed by sharing measurements in the \ac{mvf} and in the \ac{dkf}. The above challenges can be addressed by employing the \ac{btlf} approach, which allows the sharing of limited information among tracking systems under different operators and configurations.

The main contributions of our work are two-fold:  (1) We formulate a new \ac{btl}
approach for tracking a nonlinear model using the \ac{ukf} or \ac{ckf} in two separated tracking systems. (2) Our simulation results reveal new insights into the performance and numerical stability
of the \ac{ukf} and \ac{ckf} in two separated tracking systems with knowledge sharing between tracking filters. Our simulation results show significant gains in estimation accuracy of the proposed tracking framework compared to the isolated \ac{ukf} or \ac{ckf}. An interesting observation is that the \ac{ukf} integrated with \ac{btl}, referred to as \ac{tl-ukf}, is more sensitive to the value of the \ac{ukf} scaling parameter $\kappa$ (see Section~\ref{subsec:source_filter}) than the isolated \ac{ukf}. In particular, as $\kappa$ is increased, the improvement in the performance of \ac{tl-ukf} is greater than that of the isolated \ac{ukf}.  Similarly, increasing the cubature degree of the \ac{ckf} from three (equivalent to \ac{ukf} with $\kappa = 0$) to five results in a greater performance gain of \ac{tl-ckf} relative to the isolated CKF. Furthermore, integrating \ac{btl} into \ac{ukf}, third-degree \ac{ckf}, fifth-degree \ac{ckf}  outperforms not only the estimation accuracy of the isolated filters, it also performs slightly better than \ac{mvf}~\cite{Willner-1976-MVF}, which is known
to outperform other fusion approaches for multi-sensor systems such 
as state vector fusion~\cite{Roecker:1988}.

The remainder of the paper is organized as follows.  In Section~\ref{Sys_model_Section}, we provide a brief overview of the state-space representation and the traditional Bayesian filtering approach for object tracking in two separated tracking systems with mismatched intensity levels of measurement noise. Section~\ref{BTL_Section} introduces the \ac{btl} framework for tracking a nonlinear dynamic system model using two tracking filters.  Section~\ref{GM_BTL_section} derives the Gaussian process resulting from incorporating \ac{btl} into the Bayesian filtering approach under the Gaussian assumption. Section~\ref{UKF_Section} discusses the proposed tracking algorithm that integrates transfer learning into the \ac{ukf} in the source and primary filters. Section~\ref{NI_UKF_Section} addresses the numerical instability issue in the \ac{ukf}. Section~\ref{CKF_Section} addresses
integration of transfer learning into the third-degree and fifth-degree \ac{ckf}. Simulation results for the proposed algorithm are presented in Section~\ref{sect_TL_UKF_Sim}.  The paper is concluded in Section~\ref{sect_conclusion}.

\section{System Model} \label{Sys_model_Section}

\subsection{State Space Representation} \label{SSR_Section}

The discrete-time \ac{ssr} for describing a general dynamical system is given as:
\begin{align} 
\bx_k &= f_k(\bx_{k-1}) + \bv_{k-1} , \label{eq:ssr_1} \\ 
\bz_k &= h_k(\bx_k) + \bw_k , \label{eq:ssr_2}
\end{align}
where $k\in \bbN_{0}$ denotes the discrete time step. The vector $\bx_k\in\bbR^{n_\bx}$ represents the state of the system at time step $k$ with a dimension of $n_\bx$. The function $f_k: \bbR^{n_\bx} \rightarrow \bbR^{n_\bx}$ is the dynamic transition function, which describes how the state evolves over time. The vector $\bz_k \in\bbR^{n_\bz}$ represents the measurements obtained from the system at time step $k$ with a dimension of $n_\bz$. The function $h_k: \bbR^{n_\bx} \rightarrow \bbR^{n_\bz}$ is the measurement function that relates the state to the measurements. The variables $\bv_{k-1} \in \bbR^{n_{\bx}}$ and $\bw_k \in \bbR^{n_{\bz}}$ correspond to the process noise and measurement noise with dimensions of $n_\bv$ and $n_\bw$, respectively, representing the uncertainties and disturbances in the system~\cite{Ho-1964-Bayesian}.

\subsection{Bayesian Filter Approach}
\label{Bayesian_section}

In the Bayesian filter approach, the posterior \ac{pdf} of the state is constructed by incorporating all available statistical information and the sequence of observations. The extraction of the estimated state from the posterior \ac{pdf} provides an optimal solution to address the estimation problem~\cite{RistiAG-2004,Gordon-1993-Bayesian}. The estimation of the state $\bx_k$ within the Bayesian filter approach is conducted through a two-step process:

$\bullet$ {\bf Prediction Step:} The computation of the prediction for the posterior \ac{pdf} of the next state $\bx_k$ at time step $k$, given the set of measurements up to time step $k-1$, denoted as $\bz_{1:k-1}=\left\lbrace \bz_1, \ldots, \bz_{k-1}\right\rbrace $, is performed using the Chapman-Kolmogorov equation~\cite{RistiAG-2004,ArulaMGC-2002,Ho-1964-Bayesian}.  This equation provides a framework for recursively updating the PDF based on the available information up to the previous time step as follows:
\be\label{eq:Bayes_pred}
p(\bx_{k}\!\mid\!\bz_{1:k-1})\!=\!\int_{\bbR^{n_\bx}}\!p(\bx_{k}\!\mid\!\bx_{k-1})p(\bx_{k-1}\!\mid\!\bz_{1:k-1})d\bx_{k-1} ,
\ee
where $p(\bx_{k-1} \mid \bz_{1:k-1})$ is the posterior density of the state at the previous time step $k-1$ and $p(\bx_{k} \mid \bx_{k-1})$ is the transitioning \ac{pdf} characterized by \eqref{eq:ssr_1}.

$\bullet$ {\bf Update Step:} The updated posterior \ac{pdf} for $\bx_k$ given $\bz_{1:k}$ is obtained by applying Bayes' rule immediately after the new measurement at time step $k$ is observed \cite{ArulaMGC-2002}. This process incorporates the most recent measurement information into the estimation process, resulting in an improved representation of the state's posterior density as
\be \label{eq:Bayes_rule}
p(\bx_k \mid \bz_{1:k}) = \frac{p(\bz_k \mid \bx_k)\ p(\bx_{k} \mid \bz_{1:k-1})}{p(\bz_k \mid \bz_{1:k-1})} ,
\ee
\no where $p(\bz_k \mid \bx_k)$ is the measurement likelihood density function, as described in \eqref{eq:ssr_2}, at time step $k$. Furthermore, the normalization factor
$p(\bz_k \mid \bz_{1:k-1})$ is given by
\be\label{eq:Bayes_norm_meas}
p(\bz_k \mid \bz_{1:k-1}) = \int_{\bbR^{n_\bx}} p(\bz_k \mid \bx_k)\ p(\bx_{k} \mid \bz_{1:k-1})\ d\bx_{k} .
\ee

Under the assumption of Gaussian noises in the state and the measurement models described by \eqref{eq:ssr_1} and \eqref{eq:ssr_2}, the predictive density in~\eqref{eq:Bayes_pred} and the likelihood density $p(\bz_k \mid \bx_k)$ are Gaussian. This leads to the posterior density of the state being approximated as Gaussian, characterized by its mean and covariance. Consequently, the recursive nature of the Bayesian filter approach primarily revolves around the updates of means and covariances of conditional densities over time and measurements. The two main steps of the Bayesian filter approach are the prediction and update steps.

\subsection{Two Separate Tracking Systems with Mismatched Noise Intensities}
\label{Problem_Section}

\begin{figure}
		\centering
	\includegraphics[trim={0.4cm 0.42cm 0.1cm 0.35cm},clip, scale=0.35]{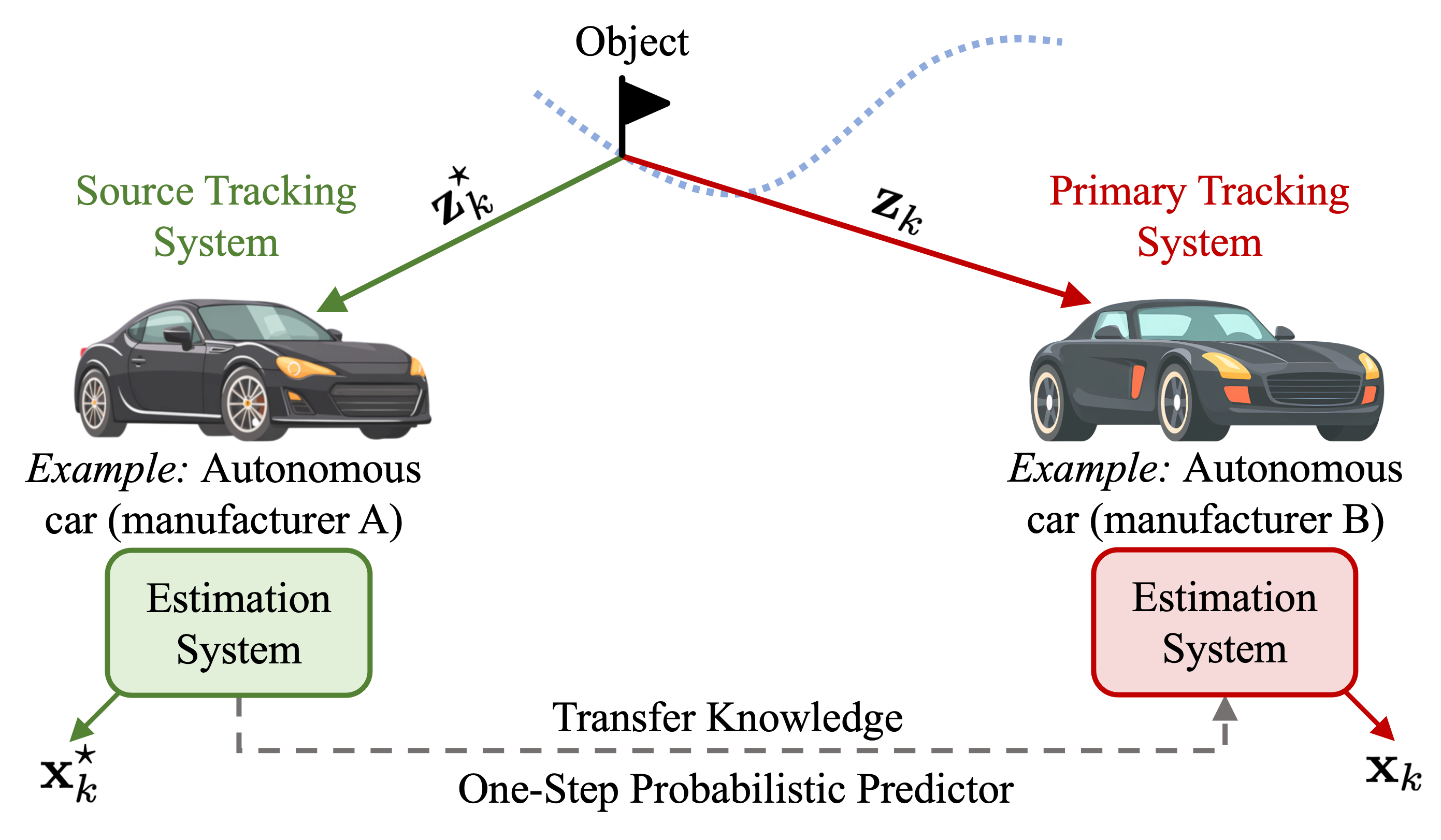}
	\caption{Graphical illustration of knowledge transfer between source and primary filters tracking the same moving object.}
	\label{fig:PR_Pred}
\end{figure}

We consider the problem of two separate tracking systems estimating an unknown state $\bx_k$ at time step $k$, assuming a state transition model defined by
\be
\bx_{k} = f\left( \bx_{k-1}\right) + \bv_{k-1} ,
\label{Pro_form_State}
\ee
where $f\left( \cdot\right)$ denotes the state transition function  and $\bv_{k-1}$ represents the error process, which is assumed to be zero-mean \ac{awgn} $\bv_{k-1}  \stackrel{\rm iid}{\sim} \ccN (\mathbf{0}, \bQ_{\bv})$ with covariance 
matrix $\bQ_\bv \in \bbR^{n_{\bx} \times n_{\bx}}$. In this paper, we consider two separate tracking systems in which
the observed measurements have mismatched noise intensities between
the two tracking systems due to various environmental or technical factors that affect the observable measurements, i.e., two autonomous vehicles built by various manufacturers or ships in maritime operating by different operators, where each has its own separated tracking system. This  leads to a reduction in estimation accuracy of the impacted tracking system. In the two tracking systems scenario, we shall denote the quantities associated
with the source tracking system with a superscript $\LR$ 
and those associated with the primary tracking system without a superscript. As illustrated in Fig.~\ref{fig:PR_Pred}, 
the measurements obtained from the sensors in both the source and primary tracking filters are affected by Gaussian noise, which are modeled by the following equations:
\begin{align}
\bz^{\LR}_{k} &= h\left( \bx_{k}\right) + \bw^{\LR}_{k}, \hspace{1cm} 
\bw^{\LR}_{k} \sim \mathcal{N}(0,\bQ^{\LR}_{\bw}), \\
\bz^{\TR}_{k} &= h\left( \bx_{k}\right) + \bw^{\TR}_{k}, \hspace{1cm} \bw^{\TR}_{k} \sim \mathcal{N}(0,\bQ^{\TR}_{\bw}),
\label{Pro_form_meas}
\end{align}
where covariance matrices for the source and primary tracking filters defined by $\bQ^{\LR}_{\bw} =I^{\LR}_{\bw}\ \bB_\bw$ and $\bQ^{\TR}_{\bw} = I_{\bw}\ \bB_\bw$, respectively, with noise intensities $I^{\LR}_{\bw}\in\bbR^{+}$ and $I_{\bw}\in\bbR^{+}$ and
common diagonal matrix $\bB_\bw \in \bbR^{n_{\bz} \times n_{\bz}}$ (see Section~\ref{subsection_Parameters_Settings} for more details). The primary tracking system is assumed to experience a higher noise intensity compared to the source tracking system, $I_{\bw}>I^{\LR}_{\bw}$. The source and primary tracking systems track the same object's state; however, each tracking system has its own estimated state, since each estimation is based on different observed measurements under different conditions. This leads to two different estimated states for the source $\bx^{\LR}_{k}$ and the primary $\bx_{k}$ tracking systems. 

\section{Bayesian Transfer Learning Filter Approach}
\label{BTL_Section}

We apply \ac{btl} to incorporate knowledge shared from the source to the primary tracking filter ~\cite{Karbalayghareh-2018-optimal_TL,Xuan-2021_bayesian_TL}. This integration of \ac{btl} within the tracking filter framework is referred to as \ac{btlf}. In particular, the source tracking filter shares its predicted observations with the primary tracking filter. This approach is applicable in scenarios where a system observes limited or unreliable measurements due to adverse conditions, such as high levels of noise intensity. 

\subsection{Source Tracking Filter}
The source filter comprises an unknown state variable set $\bx^{\LR}= \left\lbrace \bx^{\LR}_{1}, \ldots,\bx^{\LR}_{K} \right\rbrace$ and an observable measurement variable set $\bz^{\LR}= \left\lbrace\bz^{\LR}_{1}, \ldots,\bz^{\LR}_{K} \right\rbrace$. Unlike the conventional Bayesian approach, the predicted observation variable set $\bz^{\LR} = \{\bz^{\LR}_{1|0}, \ldots, \bz^{\LR}_{K+1|k} \}$ is introduced into the \ac{btlf}. The overall posterior density of the state object and predicted observation $p(\bx^{\LR}_{k}, \bz^{\LR}_{k+1|k} \mid \bz^{\LR}_{1:k})$, given the measurements set up to time step $k$, denoted as $\bz^{\LR}_{1:k}=\left\lbrace \bz^{\LR}_{1}, , \ldots, \bz^{\LR}_{k}\right\rbrace $, can be estimated by the product of two posterior densities, which are derived in Appendix~\ref{BTL_Derivation_Section}:
\be\label{eq:BTL_sr_overall}
p(\bx^{\LR}_{k}, \bz^{\LR}_{k\!+\!1|k}\!\mid\!\bz^{\LR}_{1\!:k}) \propto\  p(\bx^{\LR}_{k}\!\mid\!\bz^{\LR}_{k\!+\!1|k}, \bz^{\LR}_{1\!:k})\ p(\bz^{\LR}_{k\!+\!1|k}\!\mid\!\bz^{\LR}_{1\!:k}) .
\ee
Since, the object state $\bx^{\LR}_{k}$ at the current time step $k$ is conditionally independent of the predicted observation variable $\bz^{\LR}_{k+1|k}$ for the next time step $k+1$, the conditional probability $p(\bx^{\LR}_{k} \mid\bz^{\LR}_{k+1}, \bz^{\LR}_{1:k})$ in \eqref{eq:BTL_sr_overall} can be simplified to $p(\bx^{\LR}_{k} \mid \bz^{\LR}_{1:k})$. The posterior density of the object state $p(\bx^{\LR}_{k} \mid \bz^{\LR}_{1:k})$ can be expressed as
\be\label{eq:BTL_sr_state_1}
p(\bx^{\LR}_{k} \mid \bz^{\LR}_{1:k})\propto\, p(\bz^{\LR}_{k} \mid \bx^{\LR}_{k})\, p(\bx^{\LR}_{k} \mid \bx^{\LR}_{k-1})\, p(\bx^{\LR}_{k-1}\!\mid\!\bz^{\LR}_{1:k-1}) ,
\ee
\no where $p(\bz^{\LR}_{k} \mid \bx^{\LR}_{k})$ represents the measurement likelihood function, and $p(\bx^{\LR}_{k} \mid \bx^{\LR}_{k-1})$ denotes the predicted state for the next time step, computed using the transition \ac{pdf} given in \eqref{Pro_form_State}. Upon obtaining the estimated object state $\bx^{\LR}_{k}$ from \eqref{eq:BTL_sr_state_1}, the posterior density of the predicted observation $p(\bz^{\LR}_{k\!+\!1|k} \mid \bz^{\LR}_{1:k})$ as described in \eqref{eq:BTL_sr_overall} can be estimated by
\be\label{eq:BTL_sr_eta_1}
p(\bz^{\LR}_{k\!+\!1|k}\!\mid\!\bz^{\LR}_{1:k})\propto p(\bz^{\LR}_{k\!+\!1|k}\!\mid\!\bx^{\LR}_{k\!+\!1})p(\bx^{\LR}_{k\!+\!1}\!\mid\!\bx^{\LR}_{k})p(\bx^{\LR}_{k}\!\mid\!\bz^{\LR}_{1:k}) ,
\ee
where $p(\bz^{\LR}_{k\!+\!1|k} \mid \bx^{\LR}_{k+1})$ follows the measurement density model described in \eqref{Pro_form_meas}. Estimation of the overall posterior density of the state object and the predicted observation, denoted as $p(\bx^{\LR}_{k}, \bz^{\LR}_{k\!+\!1|k} \mid \bz^{\LR}_{1:k})$ in \eqref{eq:BTL_sr_overall}, is achieved using a three-step process:

$\bullet$ {\bf Prediction Step:} The predictive density of the object state can be computed by applying \eqref{eq:Bayes_pred} in the traditional Bayesian approach as follows:
\be\label{eq:BTL_LR_state_1}
p(\bx^{\LR}_{k}\!\mid\!\bz^{\LR}_{1:k-1})\!=\!\int_{\bbR^{n_{\bx}}}\!p(\bx^{\LR}_{k}\!\mid\!\bx^{\LR}_{k-1})p(\bx^{\LR}_{k-1}\!\mid\!\bz^{\LR}_{1:k-1}) d\bx^{\LR}_{k-1} .
\ee

$\bullet$ {\bf Update Step:} The state posterior density is updated using Bayes' rule, as described by \eqref{eq:Bayes_rule} and \eqref{eq:Bayes_norm_meas}, yielding the following expression:
\be \label{eq:BTL_LR_state_2}
p(\bx^{\LR}_{k} \mid \bz^{\LR}_{1:k}) = \frac{p(\bz^{\LR}_{k}\mid \bx^{\LR}_{k})\ p(\bx^{\LR}_{k} \mid \bz^{\LR}_{1:k-1}) }{p(\bz^{\LR}_{k} \mid \bz^{\LR}_{1:k-1}) } ,
\ee
where $p(\bz^{\LR}_{k}\!\mid\!\bz^{\LR}_{1:k-1}) $, which is the predicted observation $p(\bz^{\LR}_{k|k\!-\!1}\!\mid\!\bz^{\LR}_{1:k-1})$ of the previous time step $k\!-\!1$, is defined as
\be\label{eq:BTL_LR_state_3}
p(\bz^{\LR}_{k} \mid \bz^{\LR}_{1:k-1})  = \int_{\bbR^{n_\bx}} p(\bz^{\LR}_{k}\mid \bx^{\LR}_{k})\ p(\bx^{\LR}_{k} \mid \bz^{\LR}_{1:k-1})\ d\bx^{\LR}_{k}
\ee
The second term $p(\bz^{\LR}_{k\!+\!1|k} \mid \bz^{\LR}_{1:k})$ in \eqref{eq:BTL_sr_overall} represents the posterior density of the predicted observation, which is estimated
in the next step.

$\bullet$ {\bf Predict Observation Step:} The predicted observation density can be computed using \eqref{eq:Bayes_pred} and \eqref{eq:Bayes_norm_meas} similar to \eqref{eq:BTL_LR_state_3} as follows:
\be\label{eq:BTL_LR_eta_1}
p(\bz^{\LR}_{k\!+\!1|k}\!\mid\!\bz^{\LR}_{1:k})\!=\! \int_{\bbR^{n_\bx}}\!p(\bz^{\LR}_{k\!+\!1|k}\!\mid\!\bx^{\LR}_{k+1})p(\bx^{\LR}_{k\!+\!1}\!\mid\!\bz^{\LR}_{1:k})d\bx^{\LR}_{k\!+\!1} ,
\ee
where $p(\bx^{\LR}_{k+1} \mid \bz^{\LR}_{1:k})$ represents the predictive density at time step $k+1$, given the set of measurements up to time step $k$ computed similar to~\eqref{eq:BTL_LR_state_1}. 

\subsection{Primary Tracking Filter}
According to \ac{fpd} in~\cite{Foley-2017_fully}, the primary tracking filter has access to the probabilistic parameters of the predicted observations of the source tracking filter where the predicted observations, $\bz^{\LR}_{1:k}= \{\bz^{\LR}_{1|0}, \ldots, \bz^{\LR}_{k|k-1} \}$, from the source tracking filter, are transferred and used simultaneously as prior knowledge to estimate a set of object state variables $\mathbf{x}^{\TR}= \left\lbrace \bx^{\TR}_{1},\ldots,\bx^{\TR}_{K} \right\rbrace$ under different conditions in the primary tracking filter given a set of measurement variables $\bz^{\TR}= \left\lbrace \bz^{\TR}_{1},...,\bz^{\TR}_{K} \right\rbrace$. Note that the estimated object state $\bx_{k}$ of the primary tracking filter differs from the estimated state $\bx^{\LR}_{k}$ of the source filter even though both tracking filters track the same object due to the estimation process of each filter conditioning on different prior knowledge and observed measurements, which yields to different estimates. By applying the \ac{btlf} into the primary tracking system, the overall posterior density of the object state given the transferred predicted observations $\bz^{\LR}_{1:k}$ and the observed measurements $\bz^{\TR}_{1:k}$ up to time step $k$ is estimated as follows:
\begin{align}\label{eq:BTL_tr_overall_1}
	p(\bx^{\TR}_{k} \mid \bz^{\TR}_{1:k}, \bz^{\LR}_{1:k}) \propto\ 
	 &p( \bz^{\TR}_{k} \mid \bx^{\TR}_{k}, \bz^{\TR}_{1:k-1},  \bz^{\LR}_{1:k})\nonumber\\ 
  &p(\bx^{\TR}_{k} \mid  \bz^{\TR}_{1:k-1}, \bz^{\LR}_{1:k}) .
\end{align}

The transfer learning framework's state posterior density, denoted as $p(\bx^{\TR}_{k} \mid  \bz^{\TR}_{1:k-1}, \bz^{\LR}_{1:k})$ is computed via Bayes' rule:
\be \label{eq:BTL_tr_state_tl}
p(\bx^{\TR}_{k}\!\mid\!\bz^{\TR}_{1\!:k\!-\!1},\!\bz^{\LR}_{1:k})\!\propto\!p(\bz^{\LR}_{k|k\!-\!1}\!\mid\!\bx^{\TR}_{k},\!\bz^{\TR}_{1\!:k\!-\!1},\!\bz^{\LR}_{1:k\!-\!1}\!)p(\bx^{\TR}_{k}\!\mid\!\bx^{\TR}_{k\!-\!1}\!) .
\ee
Under the assumption that the measurement at the current time step $k$ is conditionally independent of all previous measurements, given the current state $\bx^{\TR}_{k}$, and that the measurements $\bz^{\TR}$ and the transferred predicted observations $\bz^{\LR}$ are independent, conditioned on the current state $\bx^{\TR}_{k}$, the overall state posterior can be simplified and obtained by substituting the transfer learning state posterior from \eqref{eq:BTL_tr_state_tl} into \eqref{eq:BTL_tr_overall_1}:
\be \label{eq:BTL_tr_overall_2}
p(\bx^{\TR}_{k}\!\mid\!\bz^{\TR}_{1:k}, \bz^{\LR}_{1:k}) \propto p( \bz^{\TR}_{k}\!\mid\!\bx^{\TR}_{k}) p(\bz^{\LR}_{k|k\!-\!1}\!\mid\!\bx^{\TR}_{k}) p(\bx^{\TR}_{k} \mid \bx^{\TR}_{k-1}) .
\ee

Appendix \ref{BTL_Derivation_Section} provides a proof of the estimation procedure for the overall posterior density, as described in \eqref{eq:BTL_tr_overall_2}, which involves two distinct likelihood functions: the transferred predicted observation likelihood function $p(\bz^{\LR}_{k|k\!-\!1} \mid \bx^{\TR}_k)$ and the measurement likelihood function $p(\bz^{\TR}_k \mid \bx^{\TR}_k)$. By incorporating the transferred knowledge of the predicted observation, the overall posterior density can be estimated, which leads to enhanced accuracy in the object state estimation process, through the following two-step procedure:

$\bullet$ {\bf Prediction Step:} The prediction density of object state given measurements and transferred predicted observation up to previous time step $k-1$ can be obtained,
following \eqref{eq:Bayes_pred}, as 
\begin{align} \label{eq:BTL_TR_state_2}
p & (\bx^{\TR}_{k} \mid  \bz^{\TR}_{1:k-1}, \bz^{\LR}_{1:k-1}) \nonumber \\
& =\! \int_{\bbR^{n_{\bx}}}\! p(\bx^{\TR}_{k}\!\mid\!\bx^{\TR}_{k-1})p(\bx^{\TR}_{k-1}\!\mid\!\bz^{\TR}_{1:k-1}, \bz^{\LR}_{1:k-1})d\bx^{\TR}_{k-1} .
\end{align}

$\bullet$ {\bf Update Step:} The state posterior density is updated using two likelihood functions 
\begin{align}
 p(\bx^{\TR}_{k}\!\mid\!\bz^{\TR}_{1:k\!-\!1},\!\bz^{\LR}_{1:k})&\!=\!\frac{ p(\bz^{\LR}_{k|k\!-\!1}\!\mid\!\bx^{\TR}_{k})\ p(\bx^{\TR}_{k}\!\mid\!\bz^{\TR}_{1:k-1},\!\bz^{\LR}_{1:k\!-\!1}) }{ p( \bz^{\LR}_{k|k\!-\!1}\!\mid\!\bz^{\TR}_{1:k-1}, \bz^{\LR}_{1:k-1}) } , \label{eq:BTL_TR_state_3}\\
p(\bx^{\TR}_{k} \mid  \bz^{\TR}_{1:k}, \bz^{\LR}_{1:k})&\!=\!\frac{ p( \bz^{\TR}_{k}\!\mid\!\bx^{\TR}_{k})\ p(\bx^{\TR}_{k}\!\mid\!\bz^{\TR}_{1:k-1}, \bz^{\LR}_{1:k}) }{ p( \bz^{\TR}_{k}\!\mid\!\bz^{\TR}_{1:k-1}, \bz^{\LR}_{1:k}) } , \label{eq:BTL_TR_overall_8}
\end{align}
where  the denominators in \eqref{eq:BTL_TR_state_3} and \eqref{eq:BTL_TR_overall_8},
respectively,
are defined according to \eqref{eq:Bayes_norm_meas} as
\begin{align}
p(\bz^{\LR}_{k|k\!-\!1}\!\mid& \bz^{\TR}_{1\!:k\!-\!1}\!,\!\bz^{\LR}_{1\!:k\!-\!1}\!)\!= \nonumber \\
&\int_{\bbR^{n_\bx}}\!\!\!p(\bz^{\LR}_{k|k\!-\!1}\!\mid\!\bx^{\TR}_{k}\!)p(\bx^{\TR}_{k}\!\mid\!\bz^{\TR}_{1\!:k\!-\!1}\!,\!\bz^{\LR}_{1\!:k\!-\!1}\!)d\bx^{\TR}_{k} , \label{eq:BTL_TR_norm_1}
\end{align}
\be \label{eq:BTL_TR_norm_2}
p(\bz^{\TR}_{k}\!\mid\!\bz^{\TR}_{1\!:k\!-\!1},\!\bz^{\LR}_{1\!:k})\!=\!\!\!\int_{\bbR^{n_\bx}}\!\!p(\bz^{\TR}_{k}\!\mid\!\bx^{\TR}_{k})p(\bx^{\TR}_{k}\!\mid\!\bz^{\TR}_{1\!:k\!-\!1}, \bz^{\LR}_{1\!:k})d\bx^{\TR}_{k} .
\ee

\section{Bayesian Transfer Learning Filter Approach under  Gaussian Domain}
\label{GM_BTL_section}

In the \ac{btlf} approach, the predictive density, $p(\bx_{k} \mid \bx_{k-1})$, the measurement likelihood density, $p(\bz_{k} \mid \bx_{k})$, and the transferred predicted observation likelihood density, $p(\bz^{\LR}_{k|k\!-\!1} \mid \bx^{\TR}_k)$, follow  Gaussian distributions. This is due to the assumption that the state and measurement noises in the state space models given by \eqref{eq:ssr_1} and \eqref{eq:ssr_2} are Gaussian. Therefore, this assumption leads to a Gaussian posterior density~\cite{Gordon-1993-Bayesian,Arasaratnam-2009-CKF}. Under this Gaussian assumption, the \ac{btlf} approach can be refined to recursively update the mean and covariance of the conditional densities over time, measurements, and transferred parameters.

\subsection{Source Tracking Filter}
The overall posterior of the state and the predicted observation can be estimated through three steps: prediction, update, and predict observation steps, as follows:

$\bullet$ {\bf Prediction Step:} The predictive density of the object state in \eqref{eq:BTL_LR_state_1} is represented as a Gaussian density, $p(\bx^{\LR}_{k} \mid \bz^{\LR}_{1:k-1}) = \ccN (\bx^{\LR}_{k} ; \bhx^{\LR}_{k|k-1}, \bP^{\LR}_{k| k-1})$, where $\bhx^{\LR}_{k|k-1}$ is the mean of the predictive density and $\bP^{\LR}_{k| k-1}$ is the covariance matrix. The mean of the predictive density can be computed as follows:
\be \label{eq:BTL_LR_pred_mean_1}
\bhx^{\LR}_{k|k\!-\!1}\!=\!E[\bx^{\LR}_{k}\!\mid\!\bz^{\LR}_{1\!:k\!-\!1}]\!=\!\!\int_{\bbR^{n_{\bx}}}\!\!\!f_k(\bx^{\LR}_{k\!-\!1}\!)p(\bx^{\LR}_{k\!-\!1}\!\mid\!\bz^{\LR}_{1\!:k\!-\!1}\!)d\bx^{\LR}_{k\!-\!1} .
\ee
The associated covariance matrix $\bP^{\LR}_{k| k-1}$ of the predictive Gaussian density can be obtained as
\begin{align} \label{eq:BTL_LR_pred_cov_1}
\bP^{\LR}_{k|k-1} &= E[(\bx^{\LR}_{k}- \bhx^{\LR}_{k|k-1})\ (\bx^{\LR}_{k}- \bhx^{\LR}_{k|k-1})^T \mid \bz^{\LR}_{1:k-1}]\nonumber \\
&= \int_{\bbR^{n_{\bx}}} f_k(\bx^{\LR}_{k-1})\ (f_k(\bx^{\LR}_{k-1}))^T \ p(\bx^{\LR}_{k-1} \mid \bz^{\LR}_{1:k-1})\nonumber \\
&\hspace{1.2cm}  d\bx^{\LR}_{k-1} - \bhx^{\LR}_{k|k-1}\ (\bhx^{\LR}_{k|k-1})^{T} + \bQ^{\LR}_{\bv, k-1} .
\end{align}

$\bullet$ {\bf Update Step:} 
The predictive observation $p(\bz^{\LR}_{k} \mid \bz^{\LR}_{1:k-1}) $ described in \eqref{eq:BTL_LR_state_3} follows a Gaussian distribution. The mean $\bhz^{\LR}_{k|k-1}$ and covariance $\bP^{\LR}_{\bz\bz,k| k-1}$ of this distribution can be computed via the following equations:
\begin{align} \label{eq:BTL_LR_meas_mean_1}
\bhz^{\LR}_{k|k\!-\!1}\!=\!E[\bz^{\LR}_{k}\!\mid\!\bz^{\LR}_{1\!:k\!-\!1}]\!=\!\int_{\bbR^{n_\bx}} h_k(\bx^{\LR}_{k})p(\bx^{\LR}_{k}\!\mid\!\bz^{\LR}_{1\!:k\!-\!1})d\bx^{\LR}_{k} ,
\end{align}
and
\begin{align} \label{eq:BTL_LR_meas_cov_1}
\bP^{\LR}_{\bz\bz,k| k-1}&= E[(\bz^{\LR}_{k}- \bhz^{\LR}_{k|k-1})\ (\bz^{\LR}_{k}- \bhz^{\LR}_{k|k-1})^T \mid \bz^{\LR}_{1:k-1}] \nonumber \\
&= \int_{\bbR^{n_{\bx}}} h_k(\bx^{\LR}_{k})\ (h_k(\bx^{\LR}_{k}))^T \ p(\bx^{\LR}_{k} \mid \bz^{\LR}_{1:k-1})\nonumber \\
&\hspace{1.5cm} d\bx^{\LR}_{k} - \bhz^{\LR}_{k|k-1}\ (\bhz^{\LR}_{k|k-1})^{T} + \bQ^{\LR}_{\bw, k} ,
\end{align}
where the cross-covariance between the state and the measurement, denoted as $\bP^{\LR}_{\bx\bz,k| k-1}$, can be obtained as
\begin{align} \label{eq:BTL_LR_meas_cross_cov_1}
\bP^{\LR}_{\bx\bz,k| k-1}&= E[(\bx^{\LR}_{k}- \bhx^{\LR}_{k|k-1})\ (\bz^{\LR}_{k}- \bhz^{\LR}_{k|k-1})^T \mid \bz^{\LR}_{1:k-1}] \nonumber \\
&= \int_{\bbR^{n_{\bx}}} \bx^{\LR}_{k}\ (h_k(\bx^{\LR}_{k}))^T \ p(\bx^{\LR}_{k} \mid \bz^{\LR}_{1:k-1})\ d\bx^{\LR}_{k} \nonumber \\
&\hspace{3.2cm} - \bhx^{\LR}_{k|k-1}\ (\bhz^{\LR}_{k|k-1})^{T} .
\end{align}

Once the new observed measurement arrives at time~$k$, the posterior density of the state, as defined in Equation \eqref{eq:BTL_LR_state_2}, follows a Gaussian distribution, given by $p(\bx^{\LR}_{k} \mid \bz^{\LR}_{1:k}) = \ccN (\bx^{\LR}_{k}; \bhx^{\LR}_{k|k}, \bP^{\LR}_{k| k})$. The mean and covariance of this density, denoted as $\bhx^{\LR}_{k|k}$ and $\bP^{\LR}_{k| k}$ respectively, can be calculated using the following procedure:
\begin{align}
\bhx^{\LR}_{k|k} &= \bhx^{\LR}_{k|k-1} + \bK^{\LR}_{k} (\bz^{\LR}_{k} - \bhz^{\LR}_{k|k-1}) , \label{eq:BTL_LR_upda_mean_1} \\
\bP^{\LR}_{k| k} &= \bP^{\LR}_{k| k-1} - \bK^{\LR}_{k} \ \bP^{\LR}_{\bz\bz,k| k-1} \ (\bK^{\LR}_{k} )^{T} , \label{eq:BTL_LR_upda_cov_1}
\end{align} 
where $\bK^{\LR}_{k}$ is a linear gain given by
\be \label{eq:BTL_LR_upda_gain_1}
\bK^{\LR}_{k} = \bP^{\LR}_{\bx\bz,k| k-1} \ (\bP^{\LR}_{\bz\bz,k| k-1})^{-1} .
\ee

$\bullet$ {\bf Predicted Observation Step:} The predicted observation density in \eqref{eq:BTL_LR_eta_1} is modeled as a Gaussian distribution, expressed as $p(\bz^{\LR}_{k+1|k} \mid \bz^{\LR}_{1:k})  = \ccN (\bz^{\LR}_{k+1|k}; \bhz^{\LR}_{k+1|k}, \bP^{\LR}_{\bz\bz,k+1| k})$. In this representation, the mean of the density, $\bhz^{\LR}_{k+1|k} $, and the covariance matrix, $\bP^{\LR}_{\bz\bz,k+1| k}$, are computed via 
\be \label{eq:BTL_LR_eta_mean_1}
\bhz^{\LR}_{k\!+\!1|k}\!=\!E[\bz^{\LR}_{k\!+\!1}\!\mid\!\bz^{\LR}_{1:k}]\!=\!\int_{\bbR^{n_\bx}} h_k(\bx^{\LR}_{k\!+\!1})p(\bx^{\LR}_{k\!+\!1}\!\mid\!\bz^{\LR}_{1:k})d\bx^{\LR}_{k\!+\!1} ,
\ee
\no and
\begin{align} \label{eq:BTL_LR_eta_cov_1}
\bP^{\LR}_{\bz\bz,k+1| k}&=\!E[(\bz^{\LR}_{k\!+\!1|k} - \bhz^{\LR}_{k\!+\!1|k} )\ (\bz^{\LR}_{k\!+\!1|k} - \bhz^{\LR}_{k\!+\!1|k} )^T\!\mid\!\bz^{\LR}_{1:k}] \nonumber \\
&= \int_{\bbR^{n_{\bx}}} h_k(\bx^{\LR}_{k+1})\ (h_k(\bx^{\LR}_{k+1}))^T \ p(\bx^{\LR}_{k+1} \mid \bz^{\LR}_{1:k})\nonumber \\
&\hspace{1cm} d\bx^{\LR}_{k+1} - \bhz^{\LR}_{k+1|k}\ (\bhz^{\LR}_{k+1|k})^{T} + \bQ^{\LR}_{\bw, k} ,
\end{align}
where the mean of the predictive density, $p(\bx^{\LR}_{k+1} \mid \bz^{\LR}_{1:k})$, is computed by
\be \label{eq:BTL_LR_pred_mean_2}
\bhx^{\LR}_{k+1|k}\!=\!E[\bx^{\LR}_{k+1}\!\mid\!\bz^{\LR}_{1:k}]\!=\!\int_{\bbR^{n_{\bx}}}\!f_k(\bx^{\LR}_{k})p(\bx^{\LR}_{k}\!\mid\!\bz^{\LR}_{1:k}) d\bx^{\LR}_{k} .
\ee

Note that the mean and covariance of the predicted observation density $p(\bz^{\LR}_{k+1|k} \mid \bz^{\LR}_{1:k})$ in \eqref{eq:BTL_LR_eta_mean_1} and \eqref{eq:BTL_LR_eta_cov_1} are the same parameters of the predicted observation $p(\bz^{\LR}_{k+1} \mid \bz^{\LR}_{1:k}) $ with associated mean $\bhz^{\LR}_{k+1|k}$ and covariance $\bP^{\LR}_{\bz\bz,k+1| k}$, computed in \eqref{eq:BTL_LR_meas_mean_1} and \eqref{eq:BTL_LR_meas_cov_1}. Thus, this approach does not require any additional computational process time in the source tracking filter. Upon estimating the density parameters, $(\bhz^{\LR}_{k+1|k}, \bP^{\LR}_{\bz\bz,k+1| k})$, which characterize the Gaussian distribution of the predicted observation density $p(\bz^{\LR}_{k+1|k} \mid \bz^{\LR}_{1:k})$ for time~$k+1$, these parameters are transferred to the primary tracking filter and incorporated into the tracking framework. The main objective behind the process of transferring and incorporating the predicted observation parameters is to enhance the accuracy of estimating the state density at time~$k+1$.

\subsection{Primary Tracking Filter}

Given transferred parameters and observed measurements up to time~$k$, the posterior density of the state object, $p(\bx^{\TR}_{k} \mid  \bz^{\TR}_{1:k}, \bz^{\LR}_{1:k})$, can be estimated via the following procedure:

$\bullet$ {\bf Prediction Step:} The mean, $\bhx^{\TR}_{k|k-1}$, and covariance matrix, $\bP^{\TR}_{k|k-1} $, of the predictive density in \eqref{eq:BTL_TR_state_2} can be calculated as
\begin{align} \label{eq:BTL_TR_pred_mean_1}
\hspace{-.15cm}\bhx^{\TR}_{k|k-1}\!&=\! E[\bx^{\TR}_{k} \mid \bz^{\TR}_{1:k-1}, \bz^{\LR}_{1:k-1}] \nonumber\\
\!&=\! \int_{\bbR^{n_{\bx}}}\!f_k(\bx^{\TR}_{k-1})\, p(\bx^{\TR}_{k-1}\!\mid\!\bz^{\TR}_{1:k-1}, \bz^{\LR}_{1:k-1})\, d\bx^{\TR}_{k-1} ,
\end{align}
and 
\begin{align} \label{eq:BTL_TR_pred_cov_1}
\bP^{\TR}_{k|k-1}&\!=\!E[(\bx^{\TR}_{k}- \bhx^{\TR}_{k|k-1})(\bx^{\TR}_{k}- \bhx^{\TR}_{k|k-1})^T\!\mid\!\bz^{\TR}_{1:k-1}, \bz^{\LR}_{1:k-1}] \nonumber \\
&\!=\!\int_{\bbR^{n_{\bx}}}\!f_k(\bx^{\TR}_{k\!-\!1})(f_k(\bx^{\TR}_{k\!-\!1}))^T p(\bx^{\TR}_{k\!-\!1}\!\mid\!\bz^{\TR}_{1:k\!-\!1}, \bz^{\LR}_{1:k\!-\!1})\nonumber \\ 
&\hspace{1.2cm} d\bx^{\TR}_{k-1} - \bhx^{\TR}_{k|k-1}(\bhx^{\TR}_{k|k-1})^{T} + \bQ^{\TR}_{\bv, k-1} .
\end{align}

$\bullet$ {\bf Update Step:} In the \ac{btlf} framework, the predictive density obtained in the prediction step is updated using two likelihoods, as described in \eqref{eq:BTL_tr_overall_2}. The first likelihood function employed is that of the transferred predicted observation, which is used to update the density of the object state. This update is performed by leveraging the previously estimated parameters, $(\bhz^{\LR}_{k|k-1}, \bP^{\LR}_{\bz\bz,k| k-1})$, from the source tracking filter at time~$k-1$. The mean and covariance of the predictive transferred predicted observation density in equation \eqref{eq:BTL_TR_norm_1} can be computed as follows:
\begin{align} \label{eq:BTL_TR_eta_mean_1}
\bhz^{\TR \bet}_{k|k-1} & = E[\bz^{\LR}_{k|k-1} \mid \bz^{\TR}_{1:k-1}, \bz^{\LR}_{1:k-1} ] \nonumber \\
& = \int_{\bbR^{n_\bx}} h_k(\bx^{\TR}_{k})\ \ p(\bx^{\TR}_{k} \mid \bz^{\TR}_{1:k-1}, \bz^{\LR}_{1:k-1} )\ d\bx^{\TR}_{k} ,
\end{align}
and
\begin{align} \label{eq:BTL_TR_eta_cov_1}
&\bP^{\TR \bet}_{\bz\bz,k| k\!-\!1}\!=\!E[( \bz^{\LR}_{k|k\!-\!1}\!-\!\bhz^{\TR \bet}_{k|k\!-\!1}\!)( \bz^{\LR}_{k|k\!-\!1}\!-\!\bhz^{\TR \bet}_{k|k\!-\!1}\!)^T\!\!\mid\!\bz^{\TR}_{1\!:k\!-\!1},\!\bz^{\LR}_{1\!:k\!-\!1} ] \nonumber \\
& = \int_{\bbR^{n_{\bx}}} h_k(\bx^{\TR}_{k})\ (h_k(\bx^{\TR}_{k}))^T \ p(\bx^{\TR}_{k} \mid \bz^{\TR}_{1:k-1}, \bz^{\LR}_{1:k-1} )\ d\bx^{\TR}_{k} \nonumber \\
&\hspace{2.8cm} - \bhz^{\TR \bet}_{k|k-1}\ (\bhz^{\TR \bet}_{k|k-1})^{T} + \bP^{\LR}_{\bz\bz,k| k-1} .
\end{align}
The relationship between the state and transferred parameters is characterized by the cross-covariance, expressed as
\begin{align} \label{eq:BTL_TR_eta_cros_cov_1}
\bP^{\TR \bet}_{\bx\bz,k| k-1}&\!=\!E[(\bx^{\TR}_{k}- \bhx^{\TR}_{k|k\!-\!1})( \bz^{\LR}_{k|k\!-\!1}\!-\!\bhz^{\TR \bet}_{k|k\!-\!1}\!)^T\!\mid\!\bz^{\TR}_{1:k\!-\!1}, \bz^{\LR}_{1:k\!-\!1} ] \nonumber \\
&\!=\!\int_{\bbR^{n_{\bx}}} \bx^{\TR}_{k}\ (h_k(\bx^{\TR}_{k}))^T \ p(\bx^{\TR}_{k}\!\mid\!\bz^{\TR}_{1:k-1}, \bz^{\LR}_{1:k-1} )\ d\bx^{\TR}_{k} \nonumber \\
&\hspace{3cm} - \bhx^{\TR}_{k|k-1} \ (\bhz^{\TR \bet}_{k|k-1})^{T} .
\end{align}

In the transfer learning framework, the posterior density of the state at time~$k$ given the measurements up to time step $k-1$ and the transferred parameters up to time~$k$ is assumed to follow a Gaussian distribution, denoted as $p(\bx^{\TR}_{k} \mid  \bz^{\TR}_{1:k-1}, \bz^{\LR}_{2:k}) = \ccN (\bx^{\TR}_{k}; \bhx^{\TR \bet}_{k|k-1} , \bP^{\TR \bet}_{k| k-1})$. The mean and covariance of this state posterior density in the transfer learning framework can be obtained as follows:
\begin{align}
\bhx^{\TR \bet} _{k|k-1} &= \bhx^{\TR}_{k|k-1} + \bK^{\TR \bet} _{k}\ (\bhz^{\LR}_{k|k-1} - \bhz^{\TR \bet}_{k|k-1}), \label{eq:BTL_TR_upda_mean_1} \\
\bP^{\TR \bet} _{k| k-1} &= \bP^{\TR}_{k| k-1} - \bK^{\TR \bet} _{k} \ \bP^{\TR \bet}_{\bz\bz,k| k-1} \ (\bK^{\TR \bet} _{k} )^{T} , \label{eq:BTL_TR_upda_cov_1}
\end{align}
where the linear gain of the transfer learning framework, $\bK^{\TR \bet} _{k}$, is calculated as $\bK^{\TR \bet} _{k} = \bP^{\TR \bet}_{\bx\bz,k| k-1} \ (\bP^{\TR \bet}_{\bz\bz,k| k-1})^{-1}$ .

The update process in \eqref{eq:BTL_TR_upda_mean_1} and \eqref{eq:BTL_TR_upda_cov_1}  integrates the transferred parameters into the tracking framework, refines state density estimation, and incorporates knowledge gained from the source tracking filter. After the state has been updated with the transferred predicted observation likelihood density, the measurement likelihood is incorporated. In this case, the predictive measurement density $p(\bz^{\TR}_{k} \mid \bz^{\TR}_{1:k-1}, \bz^{\LR}_{1:k})$ in equation \eqref{eq:BTL_TR_norm_2} is assumed to be a Gaussian distribution with mean and covariance can be computed as 
\begin{align} \label{eq:BTL_TR_meas_mean_1}
\bhz^{\TR}_{k|k-1} & = E[\bz^{\TR}_{k} \mid \bz^{\TR}_{1:k-1}, \bz^{\LR}_{1:k} ] \nonumber \\
& = \int_{\bbR^{n_\bx}} h_k(\bx^{\TR}_{k})\ \ p(\bx^{\TR}_{k} \mid \bz^{\TR}_{1:k-1}, \bz^{\LR}_{1:k} )\ d\bx^{\TR}_{k} ,
\end{align}
and
\begin{align} \label{eq:BTL_TR_meas_cov_1}
\bP^{\TR}_{\bz\bz,k| k-1}&= E[( \bz^{\TR}_{k}- \bhz^{\TR}_{k|k-1} )( \bz^{\TR}_{k} - \bhz^{\TR}_{k|k-1}  )^T\!\mid\!\bz^{\TR}_{1:k-1}, \bz^{\LR}_{1:k} ] \nonumber \\
&= \int_{\bbR^{n_{\bx}}} h_k(\bx^{\TR}_{k})\ (h_k(\bx^{\TR}_{k}))^T \ p(\bx^{\TR}_{k} \mid \bz^{\TR}_{1:k-1}, \bz^{\LR}_{1:k} )\nonumber \\
&\hspace{1.5cm} d\bx^{\TR}_{k} - \bhz^{\TR}_{k|k-1} \ (\bhz^{\TR}_{k|k-1} )^{T} +
\bQ^{\TR}_{\bw, k} .
\end{align}
The cross-covariance between the state and the measurement determines the relationship and dependence of the state and the measurement, defined as 
\begin{align} \label{eq:BTL_TR_meas_cross_cov_1}
\bP^{\TR}_{\bx\bz,k| k-1}&= E[ (\bx^{\TR}_{k} - \bhx^{\TR}_{k|k-1})(\bz^{\TR}_{k}- \bhz^{\TR}_{k|k-1})^T\!\mid\!\bz^{\TR}_{1:k-1}, \bz^{\LR}_{1:k} ]\nonumber \\
&= \int_{\bbR^{n_{\bx}}} \bx^{\TR}_{k}\ (h_k(\bx^{\TR}_{k}))^T \ p(\bx^{\TR}_{k} \mid \bz^{\TR}_{1:k-1}, \bz^{\LR}_{1:k} )\ d\bx^{\TR}_{k} \nonumber\\
&\hspace{3.2cm} - \bhx^{\TR \bet} _{k|k-1}\ (\bhz^{\TR}_{k|k-1})^{T} .
\end{align}
The posterior density of the overall state in \eqref{eq:BTL_tr_overall_2} can be estimated based on the newly acquired observed measurement. This density is modeled as a Gaussian distribution, described as $p(\bx^{\TR}_{k} \mid \bz^{\TR}_{1:k}, \bz^{\LR}_{2:k}) = \ccN (\bx^{\TR}_{k}; \bhx^{\TR}_{k|k} , \bP^{\TR}_{k| k})$. The mean and covariance of this overall density can be estimated as follows:
\begin{align} 
\bhx^{\TR}_{k|k} &= \bhx^{\TR \bet} _{k|k-1}  + \bK^{\TR}_{k}\ (\bz^{\TR}_{k} - \bhz^{\TR}_{k|k-1}) , \label{eq:BTL_TR_upda_mean_2}\\
\bP^{\TR}_{k| k} &= \bP^{\TR \bet} _{k| k-1} - \bK^{\TR}_{k} \ \bP^{\TR}_{\bz\bz,k| k-1} \ (\bK^{\TR}_{k} )^{T}  , \label{eq:BTL_TR_upda_cov_2}
\end{align}
where the gain is defined as $\bK^{\TR}_{k}\!=\!\bP^{\TR}_{\bx\bz,k|k\!-\!1}(\bP^{\TR}_{\bz\bz,k|k\!-\!1})^{\!-\!1}$.

After estimating the mean and covariance via \eqref{eq:BTL_TR_upda_mean_2} and \eqref{eq:BTL_TR_upda_cov_2}, the overall posterior density of the state is estimated by leveraging transferred knowledge from the source tracking filter. Incorporating \ac{btlf} approach by utilizing and leveraging transferred knowledge improves the accuracy of the estimation process, yielding more reliable and accurate estimation of the state.

\section{Transfer Learning for UKF} \label{UKF_Section}

The \ac{ukf}~\cite{JulieU-1997-UKF} uses a set of sigma points to approximate a nonlinear function, unlike the \ac{ekf} which relies on linearization. In situations where analytical Jacobians of the transition and measurement model functions are unavailable, this feature provides the \ac{ukf} an significant advantage over the \ac{ekf}~\cite{WanV-2000}.
Due to its advantages over the \ac{ekf}, the \ac{ukf} is chosen as a local approach to approximate the \ac{btlf}, described in Section~\ref{BTL_Section}, where 
knowledge is transferred from the source to the primary tracking filter and used as a prior to improve the accuracy of object state estimation. 

\subsection{Source Tracking Filter}
\label{subsec:source_filter}

By applying the \ac{ut} method~\cite{JulieU-1997-UKF}, the set of $2n_\bx+1$ sigma points are defined as
\begin{align} \label{eq:UKF_LR_sigma_1}
\ccX^{\LR}_{0, k-1|k-1} &=\bhx^{\LR}_{k-1|k-1} , \\
\ccX^{\LR}_{j, k-1|k-1} &=\bhx^{\LR}_{k-1|k-1}+ \sqrt{(n_\bx+\la)\bP^{\LR}_{j, k-1|k-1}} , \\
\ccX^{\LR}_{n_\bx+j, k-1|k-1} &=\bhx^{\LR}_{k-1|k-1}-\sqrt{(n_\bx+\la)\bP^{\LR}_{j, k-1|k-1}} ,
\end{align}
where $j=1,\ldots, n_\bx$ and $\la = \alp^2\ (n_\bx + \ka) - n_\bx$ is a scaling parameter consisting of $10^{-4} \leq \alp \leq 1$. The value of $\kappa$ is defined by $\kappa = 3 - n_\bx$, as suggested in~\cite{JulieU-1997-UKF}. Furthermore, $\bP^{\LR}_{j, k-1|k-1}$ refers to the $j$th column of the covariance matrix. The associated weights of these sigma points are obtained as
\begin{align}
W^{\LR}_{0, k}\!=\! \frac{\la}{n_\bx+\la},~~
W^{\LR}_{j, k}\!=\!\frac{1}{2n_\bx+2\la},~~ j\!=\!1,\!2,\!\ldots,\!2n_\bx . \label{eq:UKF_LR_weight_1}
\end{align}

\noindent
$\bullet$ {\bf Prediction Step:} 
The mean $\bhx^{\LR}_{k|k-1}$ and the associated covariance $\bP^{\LR}_{k|k-1}$ for the predictive density are estimated as
\begin{align}
\bhx^{\LR}_{k|k-1} &= \sum_{j=0}^{2n_\bx}\ W^{\LR}_{j,k}\ f\left( \ccX^{\LR}_{j,k-1|k-1}\right) , \label{eq:UKF_LR_pred_mean_1} \\
\bP^{\LR}_{k|k-1} &= \sum_{j=0}^{2n_\bx}\ W^{\LR}_{j,k}\ 
    \left[f\left( \ccX^{\LR}_{j,k-1|k-1}\right) - \bhx^{\LR}_{k|k-1} \right] \nonumber \\
&~~\left[f\left( \ccX^{\LR}_{j, k-1|k-1}\right) - \bhx^{\LR}_{k|k-1} \right]^{T} + \bQ^{\LR}_{\bv} . \label{eq:UKF_LR_pred_cov_1}
\end{align}
By applying the nonlinear transition model function to the sigma points, the predicted sigma points $\ccX^{\LR}_{j,k|k-1}$ can be obtained as
\be \label{eq:UKF_LR_pred_sigma_1}
\ccX^{\LR}_{j,k|k-1}= f\left( \ccX^{\LR}_{j,k-1|k-1}\right) .
\ee

\noindent
$\bullet$ {\bf Update Step:} 
The predicted measurement mean $\bhz^{\LR}_{k|k-1}$ and the associated covariance $\bP^{\LR}_{\bz\bz,k| k-1}$ can be computed as
\begin{align}
\bhz^{\LR}_{k|k-1} &= \sum_{j=0}^{2n_\bx}\ W^{\LR}_{j,k}\ h\left( \ccX^{\LR}_{j,k|k-1}\right) , \label{eq:UKF_LR_meas_mean_1} \\
\bP^{\LR}_{\bz\bz,k| k-1} &=  \sum_{j=0}^{2n_\bx}\ W^{\LR}_{j,k}\ \left[h\left( \ccX^{\LR}_{j,k|k-1}\right) - \bhz^{\LR}_{k|k-1} \right]\nonumber \\
&~~~ \left[h\left( \ccX^{\LR}_{j,k|k-1}\right) - \bhz^{\LR}_{k|k-1} \right]^{T} + \bQ^{\LR}_{\bw} .
\label{eq:UKF_LR_meas_cov_1}
\end{align}
The joint conditional density's cross-covariance can be computed as follows:
\begin{align} 
	\bP^{\LR}_{\bx\bz,k| k-1} &=  \sum_{j=0}^{2n_\bx}\ W^{\LR}_{j,k}\ \left[f\left( \ccX^{\LR}_{j,k-1|k-1}\right) - \bhx^{\LR}_{k|k-1} \right]\nonumber \\
  &~~~ \left[h\left( \ccX^{\LR}_{j,k|k-1}\right) - \bhz^{\LR}_{k|k-1} \right]^{T} .\label{eq:UKF_LR_cross_cov_1}
\end{align}
After observing the new measurement, the posterior density of the object state can be estimated with the mean $\bhx^{\LR}_{k|k} $ and covariance $\bP^{\LR}_{k|k} $,
computed as follows:
\begin{align}
\bhx^{\LR}_{k|k}  &= \bhx^{\LR}_{k|k-1}  + \bK^{\LR}_{k}  (\bz^{\LR}_{k}  - \bhz^{\LR}_{k|k-1} ) ,\label{eq:UKF_LR_upda_mean_1} \\
\bP^{\LR}_{k| k}  &= \bP^{\LR}_{k| k-1}  - \bK^{\LR}_{k} \ \bP^{\LR}_{\bz\bz,k| k-1}  \ (\bK^{\LR}_{k} )^{T} , \label{eq:UKF_LR_upda_cov_1}
\end{align}
where the Kalman gain $\bK^{\LR}_{k} $ is computed as
\be \label{eq:UKF_LR_upda_gain_1}
\bK^{\LR}_{k} = \bP^{\LR}_{\bx\bz,k| k-1} \ (\bP^{\LR}_{\bz\bz,k| k-1} )^{-1} .
\ee

\noindent
$\bullet$ {\bf Predict Observation Step:} The predicted observation variable set $\bz^{\LR}$ is introduced in the \ac{btlf} via the posterior density of the predicted observation $p(\bz^{\LR}_{k+1|k} \mid \bz^{\LR}_{1:k})$ in \eqref{eq:BTL_sr_eta_1}, where $\bz^{\LR}_{k+1|k} \sim \ccN (\bhz^{\LR}_{k+1|k}, \bP^{\LR}_{\bz\bz,k+1| k})$
is a Gaussian vector with mean $\bhz^{\LR}_{k+1|k}$ and associated covariance $\bP^{\LR}_{\bz\bz,k+1| k}$ computed as:
\begin{align}
\ccX^{\LR}_{0,k|k} &=\bhx^{\LR}_{k|k} , \label{eq:UKF_LR_sigma_2} \\
\ccX^{\LR}_{j,k|k} &=\bhx^{\LR}_{k|k}+ \sqrt{(n_\bx+\la)\bP^{\LR}_{j,k|k}} , \\
\ccX^{\LR}_{n_\bx+j,k|k} &=\bhx^{\LR}_{k|k}- \sqrt{(n_\bx+\la)\bP^{\LR}_{j,k|k}} , \\
\bhz^{\LR}_{k+1|k} &= \sum_{j=0}^{2n_\bx}\ W^{\LR}_{j,k}\ h\left( f\left( \ccX^{\LR}_{j,k|k}\right) \right) ,
\label{eq:UKF_LR_eta_mean_1} \\
\bP^{\LR}_{\bz\bz,k+1| k} &=  \sum_{j=0}^{2n_\bx}\ W^{\LR}_{j,k}\ \left[h\left( f\left( \ccX^{\LR}_{j,k|k}\right)\right)  - \bhet^{\LR}_{k+1|k}\right]\nonumber \\
 &~~~ \left[h\left( f\left( \ccX^{\LR}_{j,k|k}\right) \right)  - \bhet^{\LR}_{k+1|k}\right]^{T} + \bQ^{\LR}_{\bw} .\label{eq:UKF_LR_eta_cov_1}
\end{align}
The predicted observation parameters $\bhz^{\LR}_{k+1|k}$ and $\bP^{\LR}_{\bz\bz,k+1| k} $ are transferred to the primary tracking filter and utilized as a prior to improve estimation accuracy. 

\subsection{Primary Tracking Filter}

Similarly to the source tracking filter, the set of $2n_\bx+1$ sigma points and their associated weights in the primary tracking filter can be drawn and computed as follows:
\begin{align} \label{eq:UKF_TR_sigma_1}
	\ccX^{\TR}_{0,k-1|k-1} &=\bhx^{\TR}_{k-1|k-1} ,  \\
	\ccX^{\TR}_{j,k-1|k-1} &=\bhx^{\TR}_{k-1|k-1}+\sqrt{(n_\bx+\la)\bP^{\TR}_{j,k-1|k-1}} , \\
	\ccX^{\TR}_{n_\bx+j,k-1|k-1} &=\bhx^{\TR}_{k-1|k-1}- \sqrt{(n_\bx+\la)\bP^{\TR}_{j,k-1|k-1}} , \\
	W^{\TR}_{0,k}\!=\!\frac{\la}{n_\bx+\la}&,~~
	W^{\TR}_{j,k}\!=\!\frac{1}{2n_\bx+2\la} ,~~ j\!=\!1,\!2,\!...,\!2n_\bx \label{eq:UKF_TR_weight_1}
\end{align}

\noindent
$\bullet$ {\bf Prediction Step:} The predictive density mean $\bhx^{\TR}_{k|k-1}$ and covariance $\bP^{\TR}_{k|k-1}$ can be estimated as follows:
\begin{align} 
\bhx^{\TR}_{k|k-1} &= 
 \sum_{j=0}^{2n_\bx}\ W^{\TR}_{j,k}\ f\left( \ccX^{\TR}_{j,k-1|k-1}\right) ,
\label{eq:UKF_TR_pred_mean_1} \\
\bP^{\TR}_{k|k-1} &= \sum_{j=0}^{2n_\bx}\ W^{\TR}_{j,k}\ \left[f\left( \ccX^{\TR}_{j,k-1|k-1}\right) - \bhx^{\TR}_{k|k-1} \right]  \nonumber \\
&~~~ \left[f\left( \ccX^{\TR}_{j,k-1|k-1}\right) - \bhx^{\TR}_{k|k-1} \right]^{T} + \bQ^{\TR}_{\bv} .
\label{eq:UKF_TR_pred_cov_1}
\end{align}
The drawn sigma points in \eqref{eq:UKF_TR_sigma_1} are advanced in one time step using the nonlinear transition function as $\ccX^{\TR}_{j,k|k-1}= f( \ccX^{\TR}_{j,k-1|k-1})$.

\noindent
$\bullet$ {\bf Update Step:} The predicted object state density in the primary tracking filter is updated using two likelihood functions by incorporating transfer learning, as represented in \eqref{eq:BTL_tr_overall_2}, into the \ac{ukf} framework. In the initial stage, the transferred predicted observation likelihood function $p(\bz^{\LR}_{k|k\!-\!1}\!\mid\!\bx^{\TR}_{k})$ is employed to update the predicted object state density. This update process incorporates the transferred parameters, $\bhz^{\LR}_{k|k-1}$ and $\bP^{\LR}_{\bz\bz,k| k-1}$, which were estimated in the previous time step $k-1$ in the source filter, into the process. The predictive mean $\bhz^{\bet}_{k|k-1}$ and covariance $\bP^{\bet}_{\bz\bz,k| k-1}$ for the transferred predicted observation can be computed as
\begin{align} 
\bhz^{\bet}_{k|k-1} &= \sum_{j=0}^{2n_\bx}\ W^{\TR}_{j,k}\ h\left( \ccX^{\TR}_{j,k|k-1}\right) , 
\label{eq:UKF_TR_eta_mean_1} \\
\bP^{\bet}_{\bz\bz,k| k-1} &=  \sum_{j=0}^{2n_\bx}\ W^{\TR}_{j,k}\ \left[h\left( \ccX^{\TR}_{j,k|k-1}\right) - \bhz^{\bet}_{k|k-1} \right] , \nonumber \\
&~~~ \left[h\left( \ccX^{\TR}_{j,k|k-1}\right) - \bhz^{\bet}_{k|k-1} \right]^{T} + \bP^{\LR}_{\bz\bz,k| k-1} . \label{eq:UKF_TR_eta_cov_1}
\end{align}

Given the transferred parameter $\bz^{\LR}_{k|k\!-\!1}$, the posterior density of the state in the transfer learning framework, denoted as $p(\bx^{\TR}_{k} \mid  \bz^{\TR}_{1:k-1}, \bz^{\LR}_{2:k}) = \ccN (\bx^{\TR}_{k}; \bhx^{\TR \bet}_{k|k-1} , \bP^{\TR \bet}_{k| k-1})$, is estimated with mean $\bhx^{\TR \bet}_{k|k-1} $ and covariance $\bP^{\TR \bet}_{k| k-1}$ as 
\begin{align}
\bhx^{\TR \bet} _{k|k-1} &= \bhx^{\TR}_{k|k-1} + \bK^{\TR \bet} _{k}\ (\bhz^{\LR}_{k|k-1} - \bhz^{\TR \bet}_{k|k-1}), \label{eq:UKF_TR_upda_mean_eta_1} \\
\bP^{\TR \bet} _{k| k-1} &= \bP^{\TR}_{k| k-1} - \bK^{\TR \bet} _{k} \ \bP^{\TR \bet}_{\bz\bz,k| k-1} \ (\bK^{\TR \bet} _{k} )^{T} . \label{eq:UKF_TR_upda_cov_eta_1}
\end{align} 
The Kalman gain of the transfer learning framework $\bK^{\TR \bet}_{k} $ is given by $\bK^{\TR \bet}_{k}  = \bP^{\bet}_{\bx\bz,k| k-1}\ (\bP^{\bet}_{\bz\bz,k| k-1}  )^{-1}$ where 
\begin{align} 
\bP^{\bet}_{\bx\bz,k| k-1} &=  \sum_{j=0}^{2n_\bx}\ W^{\TR}_{j,k}\ \left[f\left( \ccX^{\TR}_{j,k-1|k-1}\right) - \bhx^{\TR}_{k|k-1} \right]\nonumber \\
&~~~ \left[h\left( \ccX^{\TR}_{j,k|k-1}\right) - \bhz^{\bet}_{k|k-1} \right]^{T} .
\label{eq:UKF_TR_cross_eta_cov_1}
\end{align}


Based on the estimated mean and covariance of the transfer learning state posterior density computed in \eqref{eq:UKF_TR_upda_mean_eta_1} and \eqref{eq:UKF_TR_upda_cov_eta_1} respectively, the set of $2n_\bx+1$ sigma points will be redrawn as
\begin{align}
	\ccX^{\TR \bet}_{0,k|k-1} &=\bhx^{\TR \bet}_{k|k-1} , \\
	\ccX^{\TR \bet}_{j,k|k-1} &=\bhx^{\TR \bet}_{k|k-1} + \sqrt{(n_\bx+\la)\bP^{\TR \bet}_{j,k| k-1} } , \\
	\ccX^{\TR \bet}_{n_\bx+j,k|k-1} &=\bhx^{\TR \bet}_{k|k-1} - \sqrt{(n_\bx+\la)\bP^{\TR \bet}_{j,k| k-1}},  \label{eq:UKF_TR_sigma_2}
\end{align}
where $j=1,2,...,n_\bx$.
The estimated measurement mean and covariance can be calculated as follows:
\begin{align}
\bhz^{\TR}_{k|k-1} &= \sum_{j=0}^{2n_\bx}\ W^{\TR}_{j,k}\ h\left( \ccX^{\TR \bet}_{j,k|k-1} \right) ,  
 \label{eq:UKF_TR_meas_mean_1} \\
\bP^{\TR}_{\bz\bz,k| k-1} &=  \sum_{j=0}^{2n_\bx}\ W^{\TR}_{j,k}\ \left[h\left( \ccX^{\TR \bet}_{j,k|k-1}\right) - \bhz^{\TR}_{k|k-1} \right] \nonumber \\
&~~~ \left[h\left(  \ccX^{\TR \bet}_{j,k|k-1}\right) - \bhz^{\TR}_{k|k-1} \right]^{T} + \bQ^{\TR}_{\bw} .
\label{eq:UKF_TR_meas_cov_1} 
\end{align}
The cross-covariance matrix, which expresses the relationship between state and measurement in the joint density, is computed as follows:
\begin{align} 
\bP^{\TR}_{\bx\bz,k| k-1} &=  \sum_{j=0}^{2n_\bx}\ W^{\TR}_{j,k}\ \left[\ccX^{\TR \bet}_{j,k|k-1}-\bhx^{\TR \bet}_{k|k-1} \right] \nonumber\\
&~~~ \left[h\left(  \ccX^{\TR \bet}_{j,k|k-1}\right) - \bhz^{\TR}_{k|k-1} \right]^{T} .
\label{eq:UKF_TR_cross_meas_cov_1}
\end{align}

Similarly to the source tracking filter, upon observing the measurement $\bz^{\TR}_{k} $, the estimated mean and covariance of the posterior density are computed as
\begin{align}
\bhx^{\TR}_{k|k}  &=\bhx^{\TR \bet}_{k|k-1} + \bK^{\TR}_{k}\  (\bz^{\TR}_{k}  - \bhz^{\TR}_{k|k-1} ) ,  \label{eq:UKF_TR_upda_mean_meas_1} \\
\bP^{\TR}_{k| k}  &= \bP^{\TR \bet}_{k| k-1}  - \bK^{\TR}_{k}\ \bP^{\TR}_{\bz\bz,k| k-1} \ (\bK^{\TR}_{k} )^{T} ,
\label{eq:UKF_TR_upda_cov_meas_1}
\end{align}
where $\bK^{\TR}_{k}$ denotes the Kalman gain defined by
\be \label{eq:UKF_TR_upda_gain_meas_1}
\bK^{\TR}_{k} = \bP^{\TR}_{\bx\bz,k| k-1}\ (\bP^{\TR}_{\bz\bz,k| k-1}  )^{-1} .
\ee
The posterior density of the state, conditioned on the observed measurement and the transferred predicted observation parameters, denoted as $p(\bx^{\TR}_{k} \mid \bz^{\TR}_{1:k}, \bz^{\LR}_{2:k}) = \ccN (\bx^{\TR}_{k}; \bhx^{\TR}_{k|k} , \bP^{\TR}_{k| k})$, given in \eqref{eq:BTL_tr_overall_2}, is modeled as a Gaussian density. The mean and covariance of this posterior density are estimated via \eqref{eq:UKF_TR_upda_mean_meas_1} and \eqref{eq:UKF_TR_upda_cov_meas_1}, respectively. Appendix \ref{TL-UKF_Alg_Section} provides an outline of the algorithms for \ac{ukf} with transfer learning in both the source and primary tracking filters. These algorithms include the specific steps and procedures for implementing the proposed approach. 

\section{Numerical Instability in the UKF}
\label{NI_UKF_Section}

The \ac{ukf} employs the \ac{ut} to approximate the underlying distribution using sigma points. These sigma points are strategically drawn to approximate a standard Gaussian distribution with a dimension of $n_{\bx}$. The process of drawing these sigma points through the \ac{ut} involves a scaling parameter $\ka$, which introduces additional flexibility for scaling higher-order moments. It is important to note that $\ka$ can be either positive or negative, but it must satisfy the condition $n_{\bx} + \ka \neq 0$. 

The moments of the sigma points differ from those of a standard Gaussian distribution. To minimize the
impact of these differences and reduce the approximation errors, it is suggested to set $n_{\bx} + \ka = 3$. This choice ensures that the moments of the sigma points align with the moments of a standard Gaussian distribution up to the fourth order. When moments are matched up to the fourth order for Gaussian cases, the accuracy of the approximation process is improved~\cite{JulieU-2000-UKF}.

In scenarios where the system dimension $n_{\bx} > 3$, the suggested value of $\ka$ in~\cite{JulieU-1997-UKF}, i.e., $\ka = 3 - n_{\bx}$, results in a negative value for $\ka$. This introduces challenges to the \ac{ukf} algorithm. First, the central sigma point $\bcX_{0}$, which represents the mean, will have a negative weight $W_{0} < 0$. This means that the central point, which is typically considered more important and carries a higher weight, is assigned a negative weight. Second, estimating the covariance matrix can become problematic as it may lose positive semi-definiteness~\cite{JulieU-1997-UKF}. This loss of positive semi-definiteness can result in numerical instability of the \ac{ukf} algorithm~\cite{Arasaratnam-2009-CKF}. These considerations highlight the need to address the implications of negative $\ka$ values on the stability and accuracy of the \ac{ukf} algorithm in high-dimensional systems.

\begin{figure}
	\centering
	\includegraphics[trim={2cm 9.5cm 2.2cm 9.5cm},clip, scale=0.48]{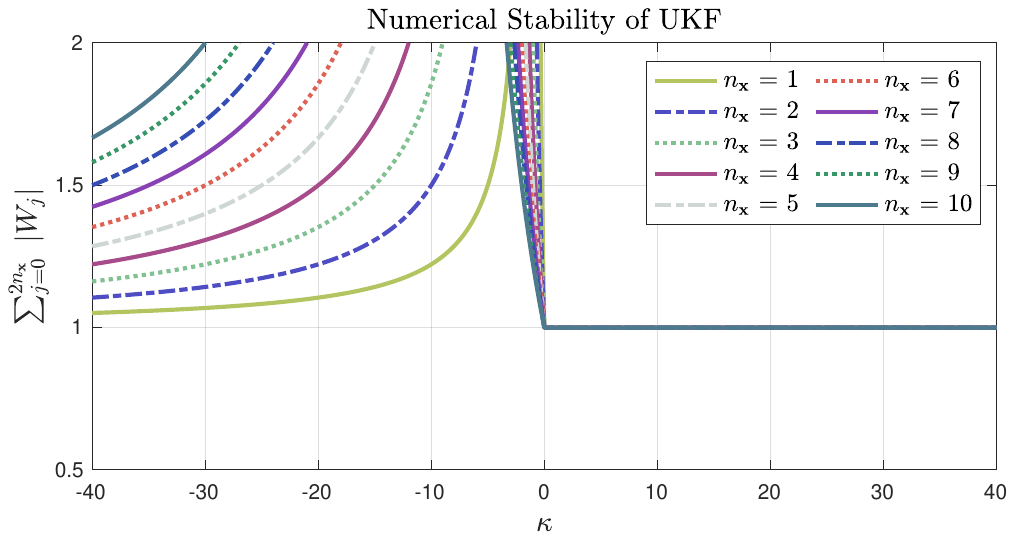}
	\caption{The stability measure addresses the impact of the $\ka$ value on the stability of the \ac{ukf} across various dimensions, $n_{\bx}$.}
	\label{fig:UKF_Stability}
\end{figure}

To address the stability of an integration rule, the absolute values of the integration rule weights must be summed to unity. This condition ensures that the integration rule is completely stable~\cite{Genz-1996-Rule_stable}. Applying this stability measure to the weights of the \ac{ukf} algorithm identifies numerical instability issues that pose challenges to the algorithm. The stability factor 
\begin{align}
	\sum_{j =0}^{2n_{\bx}} \left| W_{j} \right| &= \left| \frac{\la}{n_\bx+\la} \right| + \left| \frac{1}{2n_\bx+2\la} \right| + \cdots + \left| \frac{1}{2n_\bx+2\la} \right| \nonumber \\
	& =  \left| \frac{\la}{n_\bx+\la} \right| + 2n_\bx \left| \frac{1}{2n_\bx+2\la} \right| = \frac{n_\bx + \left| \la \right|}{\left| n_\bx + \la \right|} , \label{eq:UKF_Instability_sum_1}
\end{align}
should be unity to ensure the stability of the \ac{ukf} algorithm. 
Therefore, we must have $n_\bx + \left| \la \right| = \left| n_\bx + \la \right|$. To simplify the analysis, we assume $\alp = 1$, which implies that $\la = \ka$. Then the stability condition for the \ac{ukf}, can be expressed as $n_\bx + \left| \ka \right| = \left| n_\bx + \ka \right|$. In Fig.~\ref{fig:UKF_Stability}, the influence of different values of $\ka$ on the stability measure factor in different dimensions is examined. This demonstrates that the stability condition holds true only when $\ka \geq 0$, indicating that the \ac{ukf} is stable when $\ka$ is nonnegative.

The value of $\ka$ plays a crucial role in the numerical stability and estimation accuracy of the \ac{ukf}. Within the transfer learning framework of \ac{ukf}, the predicted observation update step in the source tracking filter, as described from \eqref{eq:UKF_LR_sigma_2} to \eqref{eq:UKF_LR_eta_cov_1}, can be influenced by the value of $\ka$.  A negative value of $\ka$ can lead to inaccurate estimates of the predicted observation density parameters that are transferred to the primary tracking filter.
This error propagation results in poor estimation accuracy at the primary tracking filter.
To mitigate numerical instability issues, the value of $\ka$ can be set to zero or a positive value for high-dimensional systems ($n_\bx > 3$). Setting $\ka$ to zero is a common solution, which effectively cancels out the central sigma point. However, the \ac{ukf} with $\ka = 0$ is identical to the third-degree \ac{ckf}. Therefore, the third-degree \ac{ckf} can be considered as a special case of the \ac{ukf}~\cite{Jia-2013-high_CKF,Chang-2012-CKF_special}. 
\section{Transfer Learning for CKF} \label{CKF_Section}

 The cubature rule was introduced as a non-product rule to solve weighted integrals that are difficult to compute, specifically Gaussian weighted integrals~\cite{Arasaratnam-2009-CKF}. The cubature rule approximates these Gaussian integrals using a set of points with associated weights, similar to the \ac{ut} approximation. As mentioned in the previous section, the third-degree \ac{ckf} is considered a special case of the \ac{ukf} with $\ka = 0$ and was introduced as a solution for the numerical instability issue that faces the \ac{ukf} when $\ka < 0$, in the case of high dimensional system. To improve the estimation accuracy of the third-degree \ac{ckf}, a fifth-degree \ac{ckf} was proposed in~\cite{Jia-2013-high_CKF}. In this section, the third-degree and fifth-degree \ac{ckf}s are integrated into the \ac{btlf}. The objective is to improve the estimation accuracy of both \ac{ckf}s by integrating transfer learning for sharing knowledge from the source to the primary tracking filter.

\subsection{Transfer Learning for Third-degree \ac{ckf}}
 The implementation of the third-degree \ac{ckf} follows a similar approach to that of the \ac{ukf} to approximate \ac{btlf}. However, the \ac{ckf}  differs from the \ac{ukf} in how the set of weighted points is drawn. Instead of using the \ac{ut}, the \ac{ckf} employs the cubature rule method. Specifically, the third-degree \ac{ckf} uses the following set of $2n_{\bx}$ sigma points:
 \begin{align}\label{eq:3rd_CKF_sigma_points_1}
\ccX_{j, k-1|k-1} &=\bhx_{k-1|k-1}+ \sqrt{n_\bx\ \bP_{j, k-1|k-1}} ,\\
\ccX_{n_\bx+j, k-1|k-1} &=\bhx_{k-1|k-1}-\sqrt{n_\bx\ \bP_{j, k-1|k-1}} ,
\end{align}
where $j=1,\ldots, n_\bx$, and the associated weights of these sigma points assigned as $W_{j, k} = \frac{1}{2n_\bx}$ for $j = 1, \ldots, n_\bx $. 

The sigma points used in the third-degree \ac{ckf} can be utilized to approximate the \ac{btlf} effectively in the source and primary tracking systems for estimating posterior densities. It is noteworthy that the associated weights for each sigma point depend only on the dimension $n_{\bx}$. This guarantees that the weights associated with the sigma points are strictly positive, thus improving the numerical stability of the third-degree \ac{ckf} relative to a \ac{ukf} with a negative value of $\ka$ (see Section~\ref{NI_UKF_Section}). Conversely, the central sigma point, which represents the mean in the  \ac{ukf}, is eliminated in \eqref{eq:3rd_CKF_sigma_points_1}. Furthermore, the distance between these chosen sigma points and the mean is proportional to the dimension $n_{\bx}$, which introduces a non-local sampling issue~\cite{JulieU-2004-UKF,Chang-2012-CKF_special}.

\subsection{Transfer Learning for Fifth-degree \ac{ckf}}

Equivalent to the third-degree \ac{ckf}, the fifth-degree of the \ac{ckf} is based on the spherical-radial cubature rule. However, it extends the degree of accuracy to the fifth degree by utilizing a set of weighted points consisting of $2n^{2}_{\bx}+1$ sigma points to compute Gaussian weighted integrals, drawn as follows:
  \begin{align}
 \ccX_{0, k-1|k-1} &=\bhx_{k-1|k-1} , \nonumber \\
\ccX_{j_a, k-1|k-1} &=\bhx_{k-1|k-1} + \gamma \sqrt{\bP_{j_a, k-1|k-1}} ,\nonumber \\
\ccX_{n_\bx+j_a, k-1|k-1} &=\bhx_{k-1|k-1} - \gamma \sqrt{\bP_{j_a, k-1|k-1}} , \nonumber \\
\ccX_{2n_\bx+j_b, k-1|k-1} &=\bhx_{k-1|k-1} +  \gamma \sqrt{\bP^{+}_{j_b, k-1|k-1}} , \nonumber 
\end{align}
\begin{align}
\ccX_{\frac{n_{\bx}(n_{\bx}+3)}{2}+j_b, k-1|k-1} &=\bhx_{k-1|k-1} - \gamma 
 \sqrt{\bP^{+}_{j_b, k-1|k-1}} , \nonumber \\
\ccX_{n_\bx(n_\bx+1)+j_b, k-1|k-1} &=\bhx_{k-1|k-1} + \gamma \sqrt{\bP^{-}_{j_b, k-1|k-1}} , \nonumber \\
\ccX_{\frac{n_{\bx}(3n_{\bx}+1)}{2}+j_b, k-1|k-1} &=\bhx_{k-1|k-1} - \gamma \sqrt{\bP^{-}_{j_b, k-1|k-1}} ,
\label{eq:5th_CKF_sigma_points_1}
 \end{align}
where $j_a = 1,\ldots, n_\bx$, $j_b = 1,\ldots, \frac{n_\bx(n_\bx-1)}{2}$,
and $\gamma = \sqrt{(n_\bx+2)}$. The matrices $\bP^{+}_{j_b, k-1|k-1}$ and $\bP^{-}_{j_b, k-1|k-1}$ in \eqref{eq:5th_CKF_sigma_points_1} are defined, respectively, as 
 \begin{align}\label{eq:5th_CKF_matrices_+_-_1}
\sqrt{\bP^{+}_{j_b, k\!-\!1|k\!-\!1}} &\!=\! 2^{-\frac{1}{2}} \left[ \sqrt{\bP_{a, k\!-\!1|k\!-\!1}} + \sqrt{\bP_{b, k\!-\!1|k\!-\!1}}\ \right] , \\
\sqrt{\bP^{-}_{j_b, k\!-\!1|k\!-\!1}} &\!=\! 2^{-\frac{1}{2}} \left[ \sqrt{\bP_{a, k\!-\!1|k\!-\!1}} - \sqrt{\bP_{b, k\!-\!1|k\!-\!1}}\ \right] ,
\end{align}
where $a<b,\ a,b = 1,\ldots, n_\bx$.
The set of $2n^{2}_{\bx}+1$ sigma points are weighted as follows:
 \begin{align}\label{eq:5th_CKF_sigma_points_weights_1}
W_{j, k} = 
\begin{cases}
\frac{2}{n_{\bx}+2} ,   &   j = 0 ,  \\
\frac{4-n_{\bx}}{2(n_{\bx}+2)^{2}} ,  &   j = 1,\ldots, 2n_\bx , \\
\frac{1}{(n_{\bx}+2)^{2}}  ,  &   j = 2n_\bx+1,\ldots, 2n^{2}_{\bx} .
\end{cases}
\end{align}

As indicated in \eqref{eq:5th_CKF_sigma_points_weights_1}, the weights associated with the fifth-degree \ac{ckf} can be negative, potentially leading to numerical instability. To address this concern, the stability measure can be applied to the weights of the fifth-degree \ac{ckf}. The sum of the absolute values of the weights is then given by
\be\label{eq:5th_CKF_Instability_sum_1}
\sum_{j =0}^{2n^{2}_{\bx}} \left| W_{j} \right| &= \frac{n_\bx \left| 4 - n_\bx \right| + 2n^{2}_{\bx} + 4 }{(n_\bx + 2)^{2}} .
\ee
According to the stability measure factor derived in \eqref{eq:5th_CKF_Instability_sum_1}, the fifth-degree \ac{ckf} achieves complete stability only when the dimension satisfies $n_\bx \leq 4$. However, in the case of high-dimensional systems where $n_\bx > 4$, the stability of the fifth-degree \ac{ckf} is less than that of the third-degree \ac{ckf}. Nevertheless, it is worth emphasizing that the central sigma point $\ccX_{0}$ of the fifth-degree \ac{ckf} will still carry a positive weight even for high-dimensional systems, which provides an advantage over the \ac{ukf} with a negative value of $\ka$, where its central sigma point has a negative weight.
\section{Simulation Results}
\label{sect_TL_UKF_Sim}

We consider the problem of estimating the unknown state vector $\bx_{k} \eqq [ x_{k}, \dot{x}_{k}, y_{k}, \dot{y}_{k}, \Omega_{k}]^T$ associated with a single object undergoing constant velocity motion in a two-dimensional trajectory. The state vector comprises the Cartesian coordinates $(x_{k}, y_{k})$, the object's velocity $(\dot{x}_{k}, \dot{y}_{k})$, and the turn rate $\Omega_{k}$. Our approach adopts a state transition model
based on a nonlinear process model employed in \cite{Shalom-2001-EAT},
which is given by
\begin{align}
\bx_{k} &= 
\begin{bmatrix}
	1 & \frac{\sin (\Omega_{k-1} T_{s})}{\Omega_{k-1}} & 0 & -\left(\frac{1-\cos (\Omega_{k-1} T_{s})}{\Omega_{k-1}}\right) & 0\\
	0 & \cos (\Omega_{k-1} T_{s}) & 0 & -\sin (\Omega_{k-1} T_{s}) & 0\\
	0 & \frac{1-\cos (\Omega_{k-1} T_{s})}{\Omega_{k-1}} & 1 &  \frac{\sin (\Omega_{k-1} T_{s})}{\Omega_{k-1}} & 0\\
	0 & \sin (\Omega_{k-1} T_{s}) & 0 & \cos (\Omega_{k-1} T_{s})  & 0\\
	0 & 0 & 0 & 0 & 1
\end{bmatrix}
\bx_{k-1} \nonumber \\
&~~~~~~+ \bv_{k-1}, 
\end{align}
where the process noise $\bv_{k-1} \sim \mathcal{N}(0,\bQ_{\bv})$ with
\be
\bQ_{\bv}=
\begin{bmatrix}
	q_{1}\frac{T_{s}^{4}}{4} & q_{1}\frac{T_{s}^{3}}{2} & 0 & 0 & 0\\
	q_{1}\frac{T_{s}^{3}}{2} & q_{1}T_{s}^{2}& 0 & 0 & 0 \\
	0 & 0 & q_{1}\frac{T_{s}^{4}}{4}& q_{1}\frac{T_{s}^{3}}{2} & 0 \\
	0 & 0 & q_{1}\frac{T_{s}^{3}}{2}  & q_{1}T_{s}^{2} & 0\\
	0 & 0 & 0 & 0 & q_{2}T_{s}
\end{bmatrix} .
\ee
In order to ensure a fair comparison, we make the assumption that both the source and primary filters possess identical error process noise covariances,
i.e., $\bQ^{\LR}_{\bv} = \bQ^{\TR}_{\bv}$.

\subsection{Parameters Settings}
\label{subsection_Parameters_Settings}
The measurement vector of the source and primary filters at time~$k$, denoted by $\bz_{k} = [ r_k, \zeta_k ]^T$, consists of the object's range $r_k$ and angle $\zeta_k$. The sensors in both the source and primary tracking filters are affected by zero-mean Gaussian noise, as given in \eqref{Pro_form_meas}, with noise intensities represented by $I^{\LR}_{\bw}$ and $I^{\TR}_{\bw}$. In both the source and primary filters, a common matrix $\bB_\bw=\text{diag}[\sigma_{r}^{2},  \sigma_{\zeta}^{2}]$  is used to ensure a fair comparison. The only differing parameters between the two tracking filters are the noise intensities. For the simulations in this section, the object follows the nonlinear transition model for a duration of $K = 100$ time steps. The initial parameters of the object state and associated covariance are set as $\bx_{0}=[1000~\mathrm{m}, 300~\mathrm{m/s}, 1000~\mathrm{m}, 0~\mathrm{m/s},  -3^{\circ}/\mathrm{s}]^{T}$ and $\bP_{0}=\text{diag}[100~\mathrm{m^2}, 10~\mathrm{m^2/s^2}, 100~\mathrm{m^2}, 10~\mathrm{m^2/s^2}, 100\times10^{-3}~\mathrm{rad^2/s^2}]$. The results obtained in this section are based on averaging $10,000$ iterations of \ac{mc} simulation, using the parameter settings as $n_\bx = 5$, $n_\bz = 2$, $T_s = 1\ \mathrm{s}$, $K=100$, $\alp = 1$, $\kappa = -2\rightarrow10$, $q_1 = 0.1\ \mathrm{m^2/s^4}$, $q_2 = 1.75\times10^{-2}\ \mathrm{rad^2/s^3}$, $\sigma_{r} = 10\ \mathrm{m}$, and $\sigma_{\zeta} = \sqrt{10}\times 10^{-3}\ \mathrm{rad}$. 

\subsection{Comparative Performance Results}
\begin{figure}
	\centering
	\includegraphics[trim={0.7cm 0.55cm 1cm 1cm},clip, scale=0.48]{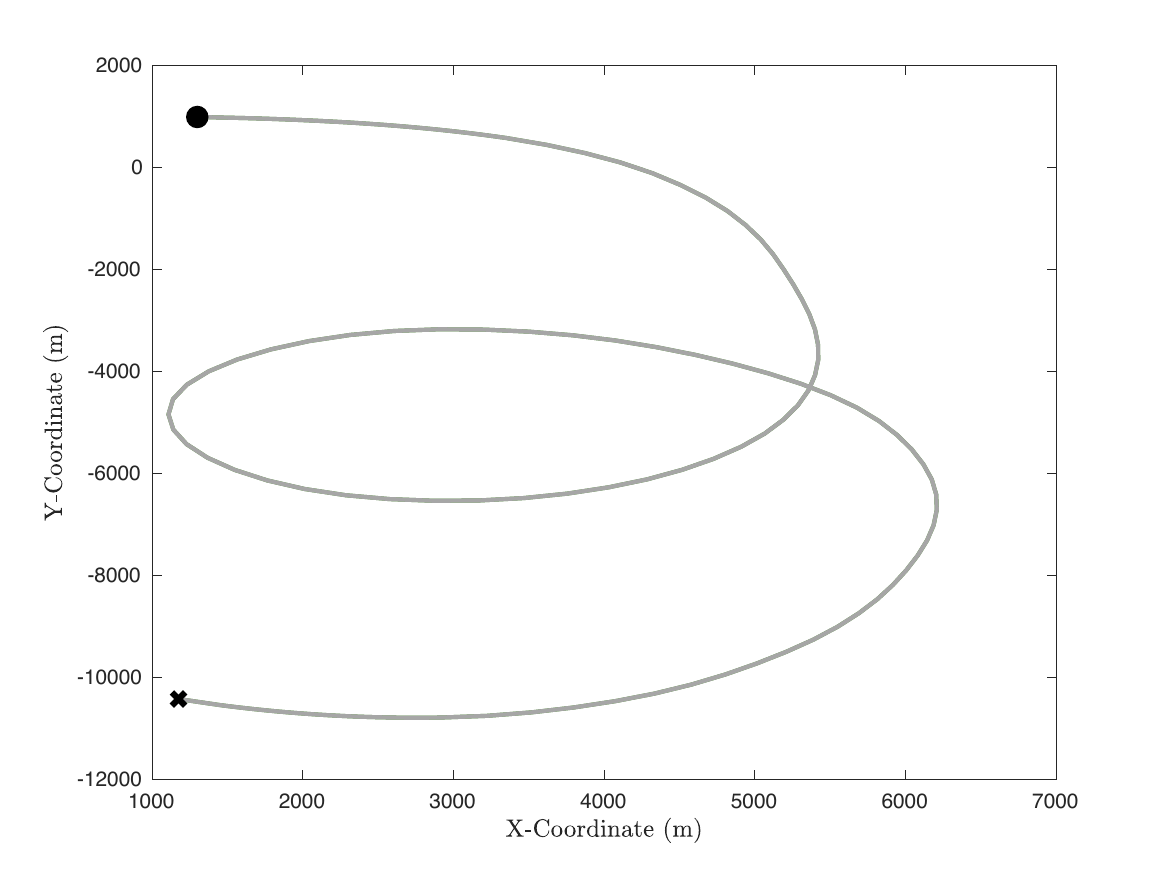}
	\caption{Object trajectory maneuvering, where $\sbullet[1.2]$ indicates the initial point of the object.}
	\label{fig:complex_traj_2}
\end{figure}

\begin{figure}
	\centering
	\includegraphics[trim={1cm 0.5cm 1cm 0.5cm},clip, scale=0.48]{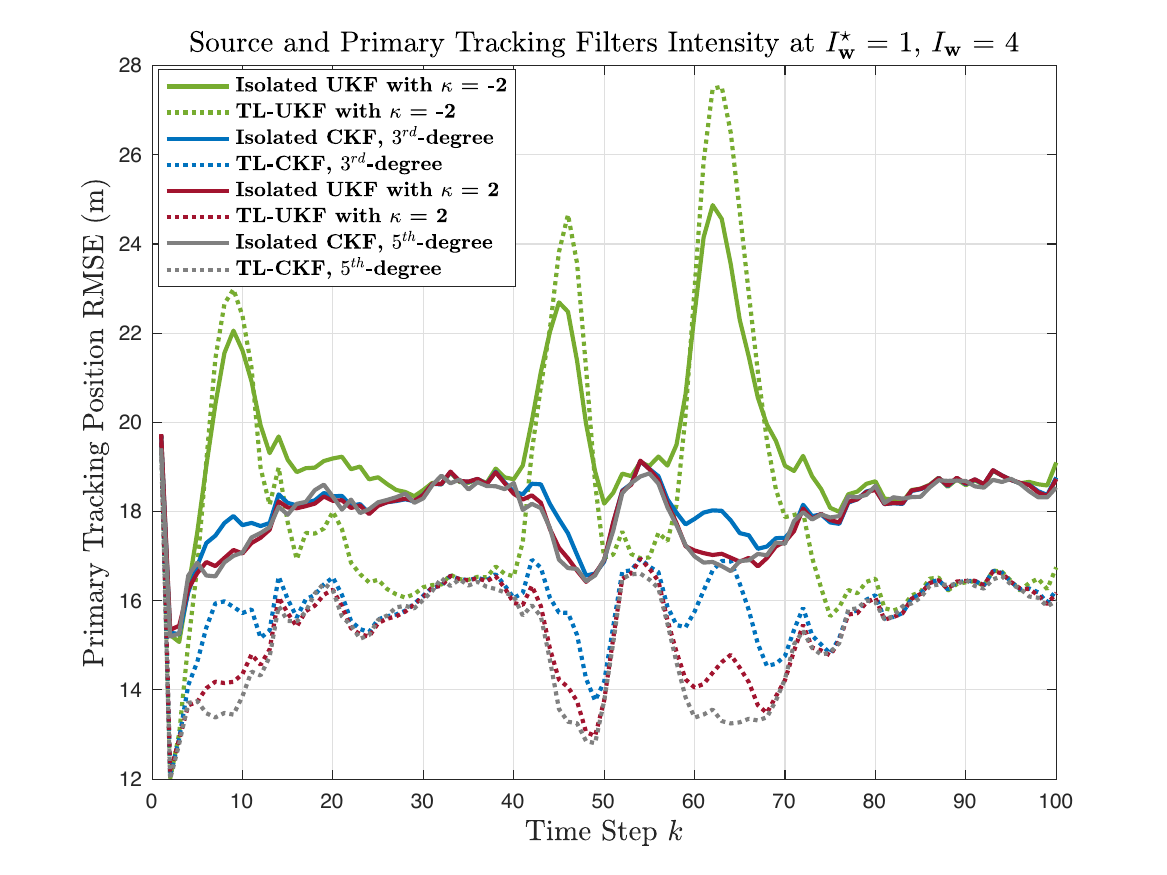}
	\caption{\ac{rmse} curves of incorporated transfer learning to the \ac{ukf} and the \ac{ckf} alongside the corresponding isolated filters under noise intensities $I^{\TR}_{\bw} = 4$ and $I^{\LR}_{\bw} = 1$.}
	\label{fig:complex_RMSE_vs_Time}
\end{figure}

We investigate the performance of the proposed \ac{tl-ukf} and \ac{tl-ckf} algorithms in a scenario involving a single maneuvering object. The trajectory of the object is depicted in Fig.~\ref{fig:complex_traj_2}. Our proposed algorithms are evaluated by computing the \ac{rmse} for the position of the object in the primary tracking filter. Under the noise intensities of $I^{\TR}_{\bw} = 4$ for the primary tracking filter and $I^{\LR}_{\bw} = 1$ for the source tracking filter, performance results in terms of \ac{rmse} for incorporating transfer learning compared to isolated filters are presented in Fig.~\ref{fig:complex_RMSE_vs_Time}. For instance, at $k=62$~s, the proposed third-degree \ac{tl-ckf}, \ac{tl-ukf} (with $\ka = 2$), and fifth-degree \ac{tl-ckf} algorithms achieve \ac{rmse} values of $16.68$~m, $14.39$~m, and $13.56$~m, respectively. In contrast, at the same time step, the isolated filters, third-degree \ac{ckf}, \ac{ukf} (with $\ka = 2$), and fifth-degree \ac{ckf}, obtain \ac{rmse} values of $18.03$~m, $17.03$~m, and $16.87$~m, respectively. As illustrated in Fig.~\ref{fig:complex_RMSE_vs_Time}, the proposed algorithms that incorporate transfer learning significantly outperform the isolated traditional filters.

However, as seen from the \ac{rmse} results in Fig.~\ref{fig:complex_RMSE_vs_Time}, the proposed \ac{tl-ukf} algorithm (with $\ka = -2$) exhibits poorer performance at certain time steps, specifically $k=\{8, 45, 62\}$~s. For example, the isolated \ac{ukf} (with $\ka = -2$), at $k=62$~s, has an \ac{rmse} of $24.87$~m, whereas the proposed \ac{tl-ukf} obtains an \ac{rmse} of $27.48$~m. This discrepancy can be attributed to the numerical instability issue, discussed in Section~\ref{NI_UKF_Section}, that arises when choosing negative values of $\ka$. The proposed \ac{tl-ukf} algorithm is significantly more sensitive to the numerical instability issue compared to the isolated \ac{ukf} due to the reliance on transfer learning, particularly in the context of predicted observations. The choice of the negative value of $\ka$ can lead to inaccurate approximations when applying the \ac{ut} via \eqref{eq:UKF_LR_sigma_2}-\eqref{eq:UKF_LR_eta_cov_1} in the predict observation step of the \ac{tl-ukf} framework. For this reason, leveraging inaccurate approximations from the source tracking filter can distort the estimation process in the primary tracking filter, yielding poorer estimation accuracy.

\subsection{Impact of Varying \texorpdfstring{$\kappa$} and Noise Intensity}
\label{subsec:varying_kappa}

\begin{figure}
	\centering
	\includegraphics[trim={1cm 0.5cm 1cm 0.5cm},clip, scale=0.48]{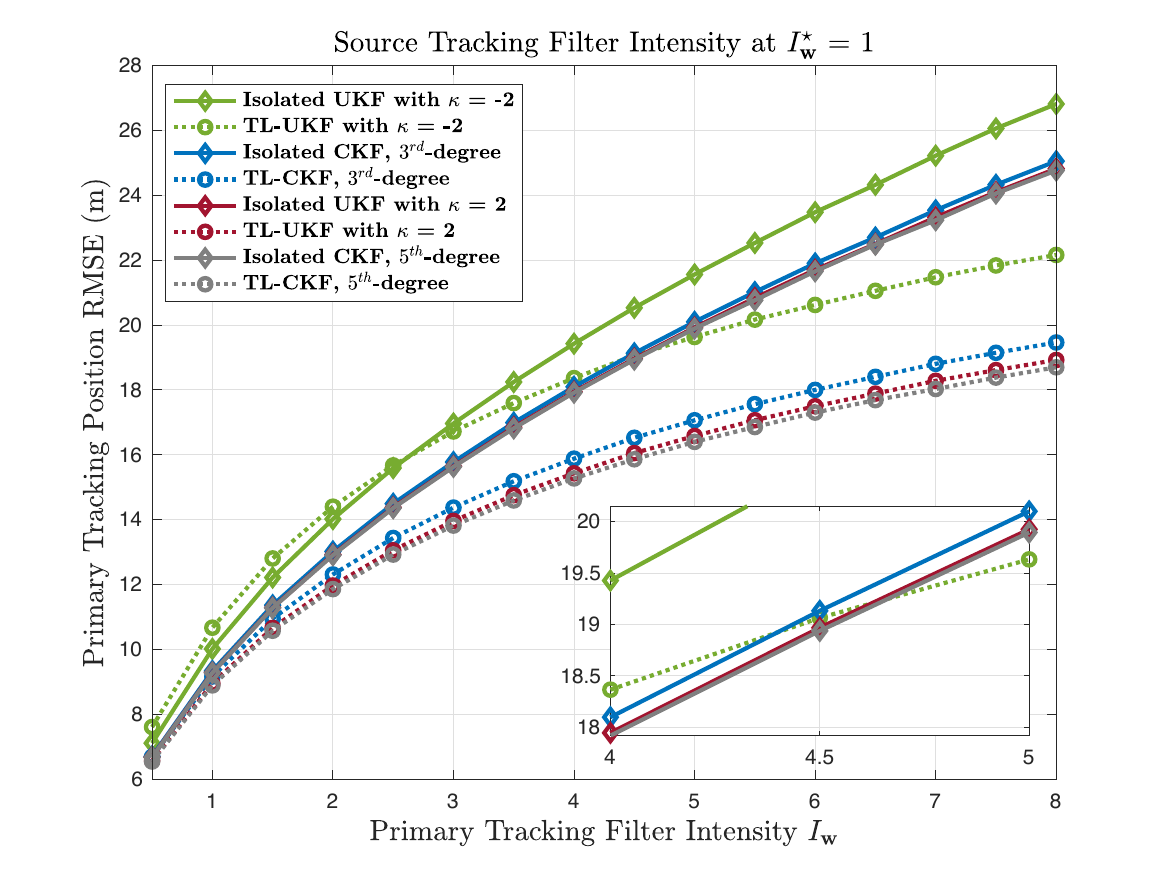}
	\caption{Performance comparison of \ac{tl-ukf}, \ac{tl-ckf}, and their isolated filter algorithms across varying levels of the primary filter noise intensity $I^{\TR}_{\bw}$.}
	\label{fig:complex_RMSE_vs_IW}
\end{figure}

The overall \ac{rmse} results under varying values of noise intensity, $I^{\TR}_{\bw} = 0.5\rightarrow8$, in the primary tracking filter when the noise intensity value of the source filter is fixed at $I^{\LR}_{\bw} = 1$ are plotted in Fig.~\ref{fig:complex_RMSE_vs_IW}. As expected, the overall performance using transfer learning (dotted lines) shows a significant improvement compared to that using isolated filters (solid lines). For instance, the proposed algorithm \ac{tl-ukf} (with $\ka = 2$) achieves an \ac{rmse} of $18.92$~m under a noise intensity value of $I^{\TR}_{\bw} = 8$, while the isolated \ac{ukf} achieves an equivalent \ac{rmse} of $18.96$~m with a lower noise intensity value of $I^{\TR}_{\bw} = 4.5$. This indicates that the proposed algorithm is capable of tracking the object with comparable estimation accuracy, even when faced with a noise intensity in the primary tracking filter that is approximately $3.5$ greater. Note that, as the noise intensity $I^{\TR}_{\bw}$ increases, the difference in terms of \ac{rmse} performance between the isolated filters (solid lines) and the proposed algorithms (dotted lines), i.e., third-degree \ac{tl-ckf}, \ac{tl-ukf} (with $\ka = 2$), and fifth-degree \ac{tl-ckf}, increases, as shown in Fig.~\ref{fig:complex_RMSE_vs_IW}.

To demonstrate the effect of the scaling parameter $\ka$ on the overall tracking performance, we simulated our proposed algorithms, a variant of the \ac{mvf} (Measurement Vector Fusion) filter, and the isolated filters with varying values of $\ka = -2 \rightarrow 10$. 
In the ideal \ac{mvf}, the measurements from the source and
primary sensors, i.e., $\bz^*$ and $\bz$, respectively, are fused according to~\cite{Willner-1976-MVF}
\begin{align}
\tilde{\bz}_k &= \bz_k + \bQ_\bw (\bQ_\bw + \bQ_\bw^*)^{-1} (\bz_k^* - \bz_k),
  \label{eq:tbz} \\
\tilde{\bQ}_\bw &= [ \bQ_\bw^{-1} + (\bQ_\bw^*)^{-1} ]^{-1} , \label{eq:tbQ}
\end{align}
where $\tilde{\bz}$ denotes the minimum mean square estimate and
$\tilde{\bQ}_w$ is the associated covariance matrix.
The fused
measurement vector $\tilde{\bz}$ 
is tracked to obtain an estimate of the state vector.
In~\cite{Roecker:1988}, the \ac{mvf} was shown to outperform state vector fusion, in which
the filtered state vectors of the two sensors are fused into a new estimate of the
state vector~\cite{Bar-Shalom:1986}.  In particular, a reduction in the covariance
of the filtered state vector is achieved by applying \ac{mvf}.
Since there is a one-step delay in transferring information
from the source to the primary filter, Eqs.~\eqref{eq:tbz} and \eqref{eq:tbQ}
cannot be implemented in practice.  Therefore, following~\cite{Foley-2017_fully},
we have implemented a variant of \ac{mvf}, in which the {\em predicted} 
observation $\boldsymbol{\eta}^*_k$ is transferred to the primary filter at time~$k$.

As seen in Table~\ref{Table:sim_2}, as the value of $\ka$ increases, the position \ac{rmse} of the primary tracking filter decreases for tracking with  transfer learning and approaches that of the 
modified \ac{mvf}. Note that the \ac{rmse} performance of \ac{tl-ukf} (with $\ka = 8$) achieves values of $8.8670$~m, $15.1862$~m, and $18.5899$~m compared to $8.8712$~m, $15.1938$~m, and $18.5959$~m when employing the \ac{mvf} approach under noise intensities $I^{\TR}_{\bw} =1$, $4$, and $8$, respectively, as given in Table~\ref{Table:sim_2}. Equivalently, the 
\ac{rmse} values of the third-degree and fifth-degree \ac{tl-ckf} are lower in comparison with those of the \ac{mvf} approach under the same conditions. The obtained \ac{rmse} values indicate that tracking with transfer learning marginally outperforms the \ac{mvf} filter.
As noted in~\cite{Foley-2017_fully}, the \ac{mvf} approach requires the strong modeling assumption
of conditional independence between $\bz^*$ and $\bz$, conditioned on a common 
$\bx$.  This assumption may be difficult to justify in applications and
is not required in the \ac{btl} approach proposed here.

\begin{table*}[htbp]
	\caption{Primary Tracking Filter Position RMSE ($\mathrm{m}$)}
        \centering
	\label{Table:sim_2}

\begin{tabularx}{\textwidth}{L{68pt}|YYY|YYY|YYY}

	 \toprule
        
	 \multirow{2}{*}{Filter} & \multicolumn{3}{c|}{$I^{\TR}_{\bw} = 1$} & %
	\multicolumn{3}{c|}{$I^{\TR}_{\bw} = 4$} & \multicolumn{3}{c}{$I^{\TR}_{\bw} = 8$} \\
        \cmidrule{2-4}  \cmidrule{5-7} \cmidrule{8-10}
        
	 & Isolated & MVF & BTLF & Isolated & MVF & BTLF & Isolated & MVF & BTLF\\
        \cmidrule{1-10}
        \rowcolor{gray!20} 
	  \ac{ukf}, $\ka = -2$ & 10.0168 & 10.6832 & 10.6702 
                             & 19.4272 & 18.3809 & 18.3650 
                             & 26.8175 & 22.1719 & 22.1621 \\

        \ac{ukf}, $\ka = 2$  & 9.2839  & 8.9564  & 8.9503 
                             & 17.9451 & 15.4398 & 15.4299 
                             & 24.8237 & 18.9328 & 18.9244 \\
	
        \rowcolor{gray!20} 
        \ac{ukf}, $\ka = 4$  & 9.2848  & 8.8942  & 8.8890
                             & 17.9283 & 15.2838 & 15.2747 
                             & 24.7914 & 18.7319 & 18.7241 \\

        \ac{ukf}, $\ka = 6$  & 9.2962  & 8.8747  & 8.8701 
                             & 17.9428 & 15.2196 & 15.2114 
                             & 24.8076 & 18.6403 & 18.6334 \\
	
        \rowcolor{gray!20} 
        \ac{ukf}, $\ka = 8$  & 9.3097  & 8.8712  & 8.8670 
                             & 17.9659 & 15.1938 & 15.1862 
                             & 24.8380 & 18.5959 & 18.5899 \\
	
        \ac{ukf}, $\ka = 10$ & 9.3230  & 8.8747  & 8.8708 
                             & 17.9909 & 15.1865 & 15.1795 
                             & 24.8721 & 18.5755 & 18.5704 \\
	
        \rowcolor{gray!20} 
   \ac{ckf}, $3^{rd}$-degree & 9.3371  & 9.1727  & 9.1651 
                             & 18.0973 & 15.8928 & 15.8815  
                             & 25.0541 & 19.4727 & 19.4637 \\
	
   \ac{ckf}, $5^{th}$-degree & 9.2784  & 8.8973  & 8.8922 
                             & 17.9178 & 15.2803 & 15.2728  
                             & 24.7675 & 18.7063 & 18.6989 \\
\bottomrule
 
\end{tabularx}

\end{table*}

\begin{figure}
	\centering
	\includegraphics[trim={1cm 0.5cm 1cm 0.5cm},clip, scale=0.48]{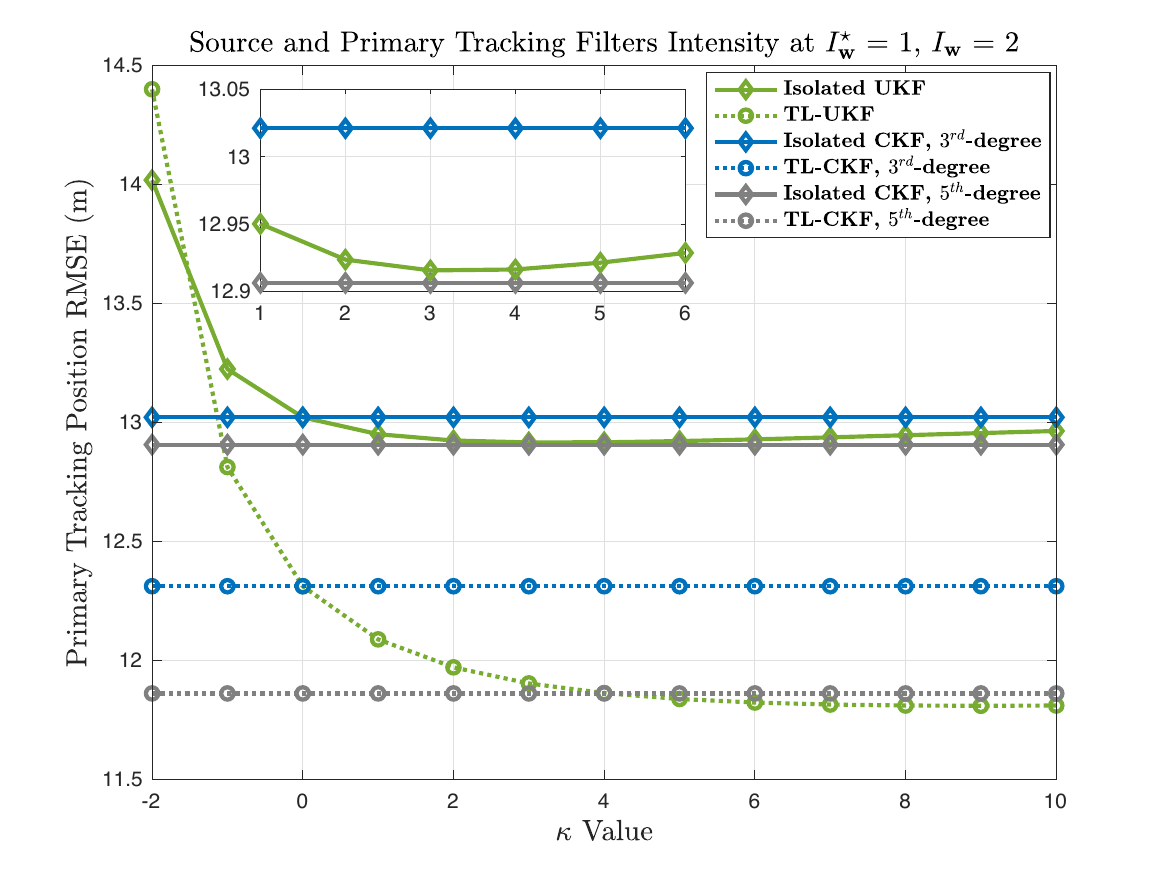}
	\caption{Performance comparison under noise intensities, $I^{\LR}_{\bw} = 1$ and $I^{\TR}_{\bw} = 2$ as a function of $\ka$ for the proposed transfer learning filters (dotted lines) and the corresponding isolated filters (solid lines).}
	\label{fig:complex_RMSE_vs_kappa}
\end{figure}

The tracking performance plotted in Fig.~\ref{fig:complex_RMSE_vs_kappa} shows
that the isolated \ac{ukf} (with $\ka > 0$) outperforms the isolated third-degree \ac{ckf}; however, the isolated \ac{ukf} does not
surpass the performance of the isolated fifth-degree \ac{ckf}. On the other hand, the proposed \ac{tl-ukf} achieves superior performance compared to the fifth-degree \ac{ckf} 
where $\ka > 4$. We observe from Fig.~\ref{fig:complex_RMSE_vs_kappa} that the \ac{rmse} performances of isolated \ac{ukf} and \ac{tl-ukf} with $\ka = 0$ are identically equal to those
of the third-degree isolated \ac{ckf} and \ac{tl-ckf}, respectively. These identical performance results are expected due to the consideration of the third-degree \ac{ckf} as a special case of the \ac{ukf} (with $\ka = 0$), as discussed in Section~\ref{NI_UKF_Section}.

\section{Conclusion} 
\label{sect_conclusion}

We proposed tracking algorithms for a pair of tracking systems that incorporate
Bayesian transfer learning into the \ac{ukf}, third-degree \ac{ckf}, and fifth-degree \ac{ckf}. Our simulation results showed a significant improvement in overall tracking accuracy of
the proposed algorithms compared to the corresponding isolated filters. Furthermore, 
by adopting the \ac{btlf} approach, a marginal improvement in overall accuracy performance compared to
a version of measurement vector fusion was obtained. This is significant because
measurement vector fusion is known to outperform 
other fusion methods in multi-sensor systems, including
state vector fusion~\cite{Roecker:1988}. On the other hand, the
\ac{mvf} approach requires a strong modeling assumption
that may not be justified in some applications (see Section~\ref{subsec:varying_kappa}).

We also investigated the impact of choosing the scaling parameter $\ka$ value for the isolated \ac{ukf} and \ac{tl-ukf}. For high-dimensional systems, with $\ka = 3 - n_{\bx}$, a negative value of $\ka$ might be selected that introduces the numerical stability issue for the \ac{ukf}. Interestingly, the \ac{tl-ukf} turns out to be highly sensitive to the value of $\kappa$, more so
than the isolated \ac{ukf}. Our simulation results showed that the performance of \ac{tl-ukf} surpasses that of the fifth-degree \ac{tl-ckf} when $\ka > 5$.
Such performance cannot be achieved with the isolated \ac{ukf}.
In this paper, the \ac{btlf} is approximated via locally approximate nonlinear filters. As future work, it would be of interest to investigate approximation of the \ac{btlf} using globally approximate approaches for tracking nonlinear trajectories.

\appendices
\section{TL-UKF Algorithms}
\label{TL-UKF_Alg_Section}

The \ac{tl-ukf} learning and tracking algorithms, based on the equations
derived in Section~\ref{UKF_Section}, are given in
Algorithms~\ref{alg_TL_UKF_LR} and~\ref{alg_TL_UKF_TR}, respectively.
Learning and tracking algorithms for the \ac{tl-ckf} based on
the equations of Section~\ref{CKF_Section}, can be formulated
similarly.

\begin{algorithm}
\caption{TL-UKF Source Tracking Algorithm}

\label{alg_TL_UKF_LR}
\begin{algorithmic}				
\State{\textbf{Inputs:}
	$\bz_{k}^{\LR}$, $\bhx_{k-1|k-1}^{\LR} $, $\bP_{k-1|k-1}^{\LR} $
	}

\State{\textbf{Outputs:}  $\bhx_{k|k}^{\LR} $, $\bP_{k|k}^{\LR} $,
  $\bhz_{k+1|k}^{\LR} $, $\bP_{\bz\bz,k+1| k}^{\LR}$
	}

\vsp{.1in}

\State{\textbf{\emph{Prediction Step}:}}

{\State Apply  \eqref{eq:UKF_LR_sigma_1} -- \eqref{eq:UKF_LR_pred_cov_1} to compute 
$\ccX_{j, k-1|k-1}^{\LR}$,
$W_{j, k}^{\LR}$,
$\bhx_{k|k-1}^{\LR}$, $\bP_{k| k-1}^{\LR}$, and
$\ccX_{j, k|k-1}^{\LR} $ for $j\eqq 0, \ldots, 2n_\bx$.}

\vsp{.05in}
\State{\textbf{\emph{Update Step}:}}

{\State Apply \eqref{eq:UKF_LR_meas_mean_1} --  \eqref{eq:UKF_LR_upda_cov_1} 
    to compute $\bhz_{k|k-1}^{\LR}$, $\bP_{\bz\bz,k| k-1}^{\LR}$, $\bP_{\bx\bz,k| k-1}^{\LR}$,
$\bK_{k}^{\LR}$,  $\bhx_{k|k}^{\LR}$, and $\bP_{k| k}^{\LR}$ .}

 \vsp{.05in}
\State{\textbf{\emph{Predicted Observation Step}:}}
{\State Apply \eqref{eq:UKF_LR_sigma_2} -- \eqref{eq:UKF_LR_eta_cov_1}
 to compute 
 $\ccX_{j, k|k}^{\LR} $,  $\ccX_{j, k+1|k}^{\LR} $,  $\bhz_{k+1|k}^{\LR}$,
 $\bP_{\bz\bz,k+1| k}^{\LR}$ for $j = 0, \ldots 2 n_\bx$.}

{\State Transfer the estimated parameters $\bhz_{k+1|k}^{\LR}$ and $\bP_{\bz\bz,k+1| k}^{\LR}$ to the primary tracking filter.}

\end{algorithmic}
\end{algorithm}

\begin{algorithm}
\caption{TL-UKF Primary Tracking Algorithm}
\label{alg_TL_UKF_TR}
\begin{algorithmic}				

\State{\textbf{Inputs:}
	 $\bz^{\TR}_{k}$,
	 $\bhx^{\TR}_{k-1|k-1} $,
  $\bP^{\TR}_{k-1|k-1}$,
  $\bhz_{k|k-1}^{\LR} $,
 $\bP_{\bz\bz,k| k-1}^{\LR}$
	}
\State{\textbf{Outputs:}  $\bhx^{\TR}_{k|k} $, $\bP^{\TR}_{k|k}$
	}

\vsp{.1in}

\State{\textbf{\emph{Prediction Step}:}}
  {\State Draw $2n_\bx +1$ sigma points  $\ccX^{\TR}_{j, k-1|k-1} $  \eqref{eq:UKF_TR_sigma_1}.}
   {\State Calculate $W^{\TR}_{j, k}$, $\bhx^{\TR}_{k|k-1}$, $\bP^{\TR}_{k| k-1}$, and
  $\ccX^{\TR}_{j, k|k-1}$ for $j\eqq 0, \ldots, 2n_\bx$ using
    \eqref{eq:UKF_TR_weight_1} -- \eqref{eq:UKF_TR_pred_cov_1}}

	\vsp{.05in}
\State{\textbf{\emph{Update Step}:}}
 {\State Calculate $\bhz^{\bet}_{k|k-1}$, $\bP^{\bet}_{\bz\bz,k| k-1} $,
      $\bP^{\bet}_{\bx\bz,k| k-1} $, $\bK_{k}^{\TR \bet}$,
      $\bhx_{k|k-1}^{\TR \bet}$, $\bP_{k| k-1}^{\TR \bet}$
      using \eqref{eq:UKF_TR_eta_mean_1} -- \eqref{eq:UKF_TR_upda_cov_eta_1}.}
      {\State Draw $2n_\bx +1$ sigma points $\ccX_{j, k|k-1}^{\TR \bet}$ \eqref{eq:UKF_TR_sigma_2}.}
      {\State Calculate $\bhz^{\TR}_{k|k-1}$, $\bP^{\TR}_{\bz\bz,k| k-1}$, $\bK^{\TR}_{k}$, $\bhx^{\TR}_{k|k}$,
        $\bP^{\TR}_{k| k}$ using \eqref{eq:UKF_TR_meas_mean_1} -- \eqref{eq:UKF_TR_upda_cov_meas_1}}
        
	\end{algorithmic}
\end{algorithm}

\section{Bayesian Transfer Learning Derivation}
\label{BTL_Derivation_Section}

\subsection{Source Tracking Filter}

The overall posterior density of the state object $\bx^{\LR}_{k}$ and the predicted observation $\bz^{\LR}_{k\!+\!1|k}$ in \eqref{eq:BTL_sr_overall} given the measurement set $\bz^{\LR}_{1:k}$ is derived by
\begin{align}\label{eq:BTL_LR_overall_1}
p(\bx^{\LR}_{k}, \bz^{\LR}_{k\!+\!1|k}\!\mid\!\bz^{\LR}_{1\!:k}) = \frac{p(\bx^{\LR}_{k}, \bz^{\LR}_{k\!+\!1|k}, \bz^{\LR}_{1:k})}{p(\bz^{\LR}_{1:k})} .
\end{align}
The joint expression $p(\bx^{\LR}_{k}, \bz^{\LR}_{k\!+\!1|k}, \bz^{\LR}_{1:k})$ in the numerator of \eqref{eq:BTL_LR_overall_1} is given by
\be \label{eq:BTL_LR_overall_2}
p(\bx^{\LR}_{k},\!\bz^{\LR}_{k\!+\!1|k},\!\bz^{\LR}_{1\!:k}\!)\!=\!p(\bx^{\LR}_{k}\!\mid\!\bz^{\LR}_{k\!+\!1|k},\!\bz^{\LR}_{1\!:k}\!) p(\bz^{\LR}_{k\!+\!1|k}\!\mid\!\bz^{\LR}_{1\!:k}\!)p(\bz^{\LR}_{1\!:k}\!) ,
\ee
where $p(\bx^{\LR}_{k}\!\mid\!\bz^{\LR}_{k\!+\!1|k}, \bz^{\LR}_{1:k})\!=\! p(\bx^{\LR}_{k}\!\mid\!\bz^{\LR}_{1:k})$, as the state posterior density at the current time step $k$ is independent of the predicted observation of the next time step $k+1$. By substituting the joint expression from \eqref{eq:BTL_LR_overall_2} into \eqref{eq:BTL_LR_overall_1}, the overall posterior density can now be rewritten as
\be\label{eq:BTL_LR_overall_3}
p(\bx^{\LR}_{k}, \bz^{\LR}_{k\!+\!1|k} \mid \bz^{\LR}_{1:k}) = p(\bx^{\LR}_{k} \mid \bz^{\LR}_{1:k})\ p(\bz^{\LR}_{k\!+\!1|k} \mid \bz^{\LR}_{1:k}) .
\ee

\subsection{Primary Tracking Filter}

The overall posterior density of the state object $\bx^{\TR}_{k}$ given the set of transferred predicated observations $\bz^{\LR}_{1:k}\!=\!\{\bz^{\LR}_{1|0}, \ldots, \bz^{\LR}_{k|k-1} \}$ and observed measurements $\bz^{\TR}_{1:k}$ up to time step $k$ is derived as
\be\label{eq:BTL_TR_overall_1}
p(\bx^{\TR}_{k} \mid  \bz^{\TR}_{1:k}, \bz^{\LR}_{1:k}) = \frac{p(\bx^{\TR}_{k}, \bz^{\TR}_{1:k}, \bz^{\LR}_{1:k})}{p( \bz^{\TR}_{1:k}, \bz^{\LR}_{1:k})} .
\ee
The numerator in \eqref{eq:BTL_TR_overall_1} can be obtained as
\begin{align}\label{eq:BTL_TR_overall_2}
\hspace{-0.15cm} p(\bx^{\TR}_{k},\! \bz^{\TR}_{1\!:k},\! \bz^{\LR}_{1\!:k})\!=&p(\bz^{\TR}_{k}\!\mid\!\bx^{\TR}_{k},\! \bz^{\TR}_{1\!:k\!-\!1},\!  \bz^{\LR}_{1\!:k})p(\bz^{\LR}_{k\!|k\!-\!1}\!\mid\!\bx^{\TR}_{k},\! \bz^{\TR}_{1\!:k\!-\!1},\! \bz^{\LR}_{1\!:k\!-\!1})\nonumber \\
& \cdot p(\bx^{\TR}_{k}\!\mid\!\bz^{\TR}_{1\!:k\!-\!1}, \bz^{\LR}_{1\!:k\!-\!1})p(\bz^{\TR}_{1\!:k\!-\!1}, \bz^{\LR}_{1\!:k\!-\!1}) .
\end{align}

Under the assumption that the measurements and the transferred predicted observations are conditionally independent given the object state $\bx^{\TR}_{k}$, the joint density in \eqref{eq:BTL_TR_overall_2} can be simplified as
\begin{align} \label{eq:BTL_TR_overall_3}
p & (\bx^{\TR}_{k}, \bz^{\TR}_{1:k}, \bz^{\LR}_{1:k}) = p( \bz^{\TR}_{k} \mid \bx^{\TR}_{k})\ p(\bz^{\LR}_{k\!|k\!-\!1} \mid \bx^{\TR}_{k})\nonumber \\
& \cdot  p(\bx^{\TR}_{k} \mid  \bz^{\TR}_{1:k-1}, \bz^{\LR}_{1:k-1})\ p(\bz^{\TR}_{1:k-1}, \bz^{\LR}_{1:k-1}) ,
\end{align}
and the denominator in \eqref{eq:BTL_TR_overall_1} can be factorized as
\begin{align}\label{eq:BTL_TR_overall_4}
&\hspace{-0.29cm} p( \bz^{\TR}_{1:k}, \bz^{\LR}_{1:k}) = p( \bz^{\TR}_{k}, \bz^{\TR}_{1:k-1}, \bz^{\LR}_{1:k})\!=\! \nonumber \\
&\hspace{-0.18cm}p(\bz^{\TR}_{k}\!\mid\!\bz^{\TR}_{1\!:k\!-\!1},\!\bz^{\LR}_{1\!:k})p( \bz^{\LR}_{k\!|k\!-\!1}\!\mid\!\bz^{\TR}_{1\!:k\!-\!1},\!\bz^{\LR}_{1\!:k\!-\!1})p( \bz^{\TR}_{1\!:k\!-\!1},\!\bz^{\LR}_{1\!:k\!-\!1}) .
\end{align}

The overall posterior density can be obtained by substituting the simplified numerator from \eqref{eq:BTL_TR_overall_3} and the factorized denominator from \eqref{eq:BTL_TR_overall_4} into \eqref{eq:BTL_TR_overall_1} as 
\begin{align}
p(\bx^{\TR}_{k}\!\mid\!\bz^{\TR}_{1\!:k},\!\bz^{\LR}_{1\!:k})&\!=\!\frac{ p( \bz^{\TR}_{k}\!\mid\!\bx^{\TR}_{k})p(\bz^{\LR}_{k\!|k\!-\!1}\!\mid\!\bx^{\TR}_{k})p(\bx^{\TR}_{k}\!\mid\!\bz^{\TR}_{1\!:k\!-\!1},\!\bz^{\LR}_{1\!:k\!-\!1}) } { p( \bz^{\TR}_{k}\!\mid\!\bz^{\TR}_{1\!:k\!-\!1},\!\bz^{\LR}_{1\!:k})p( \bz^{\LR}_{k\!|k\!-\!1}\!\mid\!\bz^{\TR}_{1\!:k\!-\!1},\!\bz^{\LR}_{1\!:k\!-\!1}) } \nonumber \\
&\!=\!\frac{ p( \bz^{\TR}_{k} \mid \bx^{\TR}_{k})\ p(\bx^{\TR}_{k} \mid  \bz^{\TR}_{1:k-1}, \bz^{\LR}_{1:k}) }{ p( \bz^{\TR}_{k} \mid \bz^{\TR}_{1:k-1}, \bz^{\LR}_{1:k}) } ,
\label{eq:BTL_TR_overall_7}
\end{align}
where the second term in the numerator of \eqref{eq:BTL_TR_overall_7} is referred to the transfer learning state posterior density and is given by
\be\label{eq:BTL_TR_state_1}
p(\bx^{\TR}_{k}\!\mid\!\bz^{\TR}_{1:k-1}, \bz^{\LR}_{1:k})\!=\!\frac{ p(\bz^{\LR}_{k\!|k\!-\!1}\!\mid\!\bx^{\TR}_{k})\ p(\bx^{\TR}_{k}\!\mid\!\bz^{\TR}_{1:k-1}, \bz^{\LR}_{1:k-1}) }{ p( \bz^{\LR}_{k\!|k\!-\!1}\!\mid\!\bz^{\TR}_{1:k-1}, \bz^{\LR}_{1:k-1}) } .
\ee

\bibliographystyle{IEEEtran}
\bibliography{IEEEabrv,library}

\end{document}